\documentclass[12pt]{iopart}

\usepackage{amsfonts}
\usepackage{amssymb}
\usepackage{iopams}
\usepackage{setstack}
\usepackage{graphicx}
\usepackage{color}
\usepackage{cite}

\def\ri{{\rm i}}
\def\vareps{\varepsilon}
\def\sgn{{\rm sgn}}

\begin{document}

\title{Asymptotic reductions and solitons of nonlocal nonlinear Schr\"{o}dinger equations}

\author{Theodoros P. Horikis$^{1}$, Dimitrios J. Frantzeskakis$^{2}$}
\address{$^{1}$Department of Mathematics, University of Ioannina, Ioannina 45110, Greece}

\address{$^{2}$Department of Physics, National and Kapodistrian University of Athens, Panepistimiopolis,
Zografos, Athens 15784, Greece}

\ead{horikis@uoi.gr}

\begin{abstract}
Asymptotic reductions of a defocusing nonlocal nonlinear Schr\"{o}dinger model in
$(3+1)$-dimensions, in both Cartesian and cylindrical geometry, are presented. First, at an
intermediate stage, a Boussinesq equation is derived, and then its far-field, in the form of a
variety of Kadomtsev-Petviashvilli (KP) equations for right- and left-going waves, is found. KP
models include versions of the KP-I and KP-II equations, in Cartesian and cylindrical geometry.
Solitary waves solutions, planar or ring-shaped, and of dark or anti-dark type, are also
predicted to occur. Their nature and stability is determined by a parameter defined by the
physical parameters of the original nonlocal system. It is thus found that (dark) anti-dark
solitary waves are only supported by a weak (strong) nonlocality, and are unstable (stable) in
higher-dimensions. Our analytical predictions are corroborated by direct numerical simulations.
\end{abstract}

\submitto{\JPA}

\section{Introduction}

In the theory of nonlinear waves, many problems appear where several different temporal and/or
spatial scales are present. Thus, asymptotic multiscale expansion methods are usually applied to
derive nonlinear evolution equations more manageable to the problem at hand \cite{Jeffrey}. By
means of such asymptotic methods, it has been demonstrated that several systems integrable by the
Inverse Scattering Transform (IST) \cite{BlackBook} can be reduced to other integrable models
\cite{zakh}. Such connections have been proved extremely useful since solutions of the reduced
models can be used to construct approximate solutions of the original models. A prominent example
is the connection of the defocusing nonlinear Schr\"{o}dinger (NLS) equation with the Korteweg-de
Vries (KdV) equation, which allowed for the description of shallow dark solitons of the former in
terms of KdV solitons. Relevant studies started in the early 70s \cite{tsu} and continue to this
date \cite{djfth}; importantly, they have also been extended to the case of non-integrable systems,
providing extremely useful information on the existence, stability and dynamics of solutions in
various physical settings, such as nonlinear optics \cite{ofy} and Bose-Einstein condensates
\cite{djf,siam}.

Indeed, these methods become even more useful when the original system is not integrable without
known solutions in explicit form. Models featuring a spatially nonlocal nonlinear response are
important examples, since they are used to describe beam dynamics and solitons in various systems
such as plasmas \cite{litvak}, atomic vapors \cite{vap}, lead glasses featuring strong thermal
nonlinearity \cite{rot}, as well as media with a long-range inter-particle interaction. The latter
include nematic liquid crystals with long-range molecular reorientational interactions \cite{ass0},
as well as dipolar bosonic quantum gases \cite{santos}. Importantly, nonlocality plays a key role
on the soliton properties. In particular, in the case of focusing nonlocal nonlinearities, collapse
does not take place in higher-dimensions \cite{tursolo,krol1}, which results in stable solitons, as
observed in experiments of, say \cite{vap,rot}, even in the $(3+1)$-dimensional setting \cite{Mih2}
-- see, e.g., reviews \cite{krol1,Mih} and references therein. On the other hand, in the case of
defocusing nonlocal nonlinearities, dark solitons that are supported in such settings
\cite{dr1,kart1,darknem,tph}, may exhibit an attractive interaction \cite{dr1}, rather than a
repulsive one, as is the situation in the case of a local nonlinearity -- cf. \cite{ofy,djf,siam}
and references therein. Furthermore, dark solitons which are known to be prone to the transverse
(or ``snaking'') instability in higher-dimensional settings \cite{kuz2,peli1,pelirev,siam}, can be
stabilized due to the nonlocal nonlinearity \cite{trillo}.

Motivated by the above, and particularly by the fact that nonlocal nonlinearities have a profound
effect on the form and stability properties of nonlinear excitations, in this work, we study a
nonlocal defocusing NLS model in $(3+1)$-dimensions. A brief description of our findings, as well
as the outline of the presentation, is as follows.

First, in Section~2, we present the model which
is considered in both physically relevant geometries: Cartesian and cylindrical. The model
describes light beam propagation in nematic liquid crystals \cite{ass1,ass2} and thermal nonlinear
media \cite{krol1}, and has been used to predict three-dimensional (3D) solitons \cite{Mih2,Mih}.
In addition, the continuous-wave (cw) solution is found and its stability is analysed.
Then, in Section~3, we use multiscale expansion methods to derive asymptotic reductions of the nonlocal NLS
equation. At an intermediate stage of the asymptotic analysis, we obtain a Boussinesq-type
equation. Next, we consider the far-field of the Boussinesq equation, and derive various
Kadomtsev-Petviashvilli (KP) type models for right- and left-going waves. These models, which are
classified in Section~4, include KP-I and KP-II equations, in Cartesian and cylindrical geometry
(in the latter, the relevant model is also known as the Johnson equation \cite{J1}), as well as the
Cylindrical-I (CI) and Cylindrical-II (CII) equations. All these models have been used to describe
various physical situations ranging from shallow-water waves to ion-acoustic waves in plasmas, and
so on \cite{Karpman,Infeld,Johnson,MJA1}.
%Although particular solutions of the reduced systems are not the centerpiece here,
%
In the same Section (Sec.~4), we present solitary waves of the original nonlocal
model, which are constructed from solutions of the effective asymptotic equations. We thus find
KdV-type, spatial dark or anti-dark solitary waves, namely intensity dips or humps on top of the
cw-background, which may exhibit either a planar shape (stripes) or a cylindrical (ring) shape.
Dark (anti-dark) solitary waves exist in a weak (strong) nonlocal regime, defined by the sign of a
characteristic quantity that links all the physical parameters of the original nonlocal system, and
are unstable (stable) in the 2D setting. We also find temporal dark solitary waves which are
predicted to be unstable in higher-dimensions.
Our analytical predictions are found to be in very good agreement with results
from direct numerical simulations: indeed, using the analytical forms of the spatial
soliton solutions as initial conditions for the direct numerical integration of the
original problem, we find that the solitons propagate undistorted -- at least for
relatively small propagation distances. Note that, even for longer propagation distances,
instabilities are not observed in our simulations, which suggests that the solitons
found have a good chance to be observed in experiments.
Finally, in Section~5, we present our conclusions
and highlight some interesting directions for future studies.

\section{Model and continuous wave solution}

We consider a $(3+1)$-dimensional nonlocal NLS model,
composed by a system of two coupled equations for the complex field amplitude $u$ and the
nonlinear correction to the refractive index $n$ (which is a real function).
This model applies, in particular, to light propagation in nematic liquid crystals
\cite{ass0,ass1}
and to thermal nonlinear optical media \cite{krol1} (see also relevant theoretical
and experimental work in~\cite{krol2}). The system under consideration is expressed
in the following dimensionless form \cite{Mih2,Mih}:
\begin{eqnarray}
\ri u_{z} + \frac{1}{2} (\Delta u- Du_{tt}) -2nu =0,
\label{NLS1} \\
d\Delta n - 2qn + 2|u|^2 = 0,
\label{NLS2}
\end{eqnarray}
where subscripts denote partial derivatives,
$z$ is the propagation coordinate (normalized to the diffraction length),
$t$ is retarded time, and $\Delta$ is the transverse Laplacian. Below, we
consider both physically relevant geometries, Cartesian
and cylindrical, for which the Laplacian respectively reads:
\begin{equation*}
\Delta = \partial_x^2+\partial_y^2, \qquad
\Delta = \frac{1}{r}\partial_r(r\partial_r)+\frac{1}{r^2}\partial_{\theta}^2,
\end{equation*}
with transverse coordinates $\boldsymbol{r}_{\perp} =(x,~y)$ or
$\boldsymbol{r}_{\perp}=(r,~\theta)$, respectively, normalized with respect
to the beam width. Additionally, the real
constants $D$, $d$, and $q$ in Eqs.~(\ref{NLS1})-(\ref{NLS2}), which are assumed to be
$\mathcal{O}(1)$ parameters in our analysis below, have the following physical significance. First,
$D$ represents the ratio of diffraction and dispersion lengths,
%\tph{
with $D>0$ ($D<0$) corresponding to anomalous (normal) group-velocity dispersion (GVD).
Second, in the context of nematic liquid crystals, the parameter $q$ is related to the square
of the applied static electric field that pre-tilts the nematic dielectric \cite{pec,alb,ass2}.
Third, parameter $d$ measures the
relative width of the response of the medium to the light field, and is connected to the
nonlocality scale of the nonlinear response of the medium: in the limit $d \rightarrow 0$, the
system of Eqs.~(\ref{NLS1})-(\ref{NLS2}) decouples and is reduced to a local $(3+1)$-dimensional
NLS equation with a cubic defocusing nonlinearity.

To start our analysis, we use the Madelung transformation
\begin{equation}
u=u_0\sqrt{\rho}\exp(i\phi),
\label{Mand}
\end{equation}
($u_0$ being an arbitrary complex constant) to separate the real functions for the
amplitude $\rho$ and phase $\phi$ of $u$
in Eq.~(\ref{NLS1}), and derive from Eqs.~(\ref{NLS1})-(\ref{NLS2})
the following system:
\begin{eqnarray}
\phi_z + 2n +\frac{1}{2}\left[\left(\boldsymbol{\nabla} \phi\right)^2-D\phi_t^2\right]
-\frac{1}{2}\rho^{-1/2}\left[\Delta \rho^{1/2} -D\left(\rho^{1/2}\right)_{tt}\right]=0,
\label{h1} \\
\rho_z+ \boldsymbol{\nabla} \cdot (\rho \boldsymbol{\nabla}\phi) - D (\rho \phi_t)_t =0,
\label{h2} \\
d\Delta n -2qn+2|u_0|^2\rho=0,
\label{h3}
\end{eqnarray}
where $\boldsymbol{\nabla}$ is the typical gradient vector operator, defined as:
\[
\boldsymbol{\nabla} =(\partial_x,~\partial_y), \qquad
\boldsymbol{\nabla} =\left(\partial_r,~\frac{1}{r}\partial_{\theta}\right)
\]
for the Cartesian or the cylindrical geometry, respectively.

It is readily observed that the above system possesses an exact steady-state solution
of the form:
\[
\phi = - \frac{2}{q}|u_0|^2 z, \qquad \rho = 1, \qquad  n = \frac{1}{q}|u_0|^2,
\]
which corresponds to the continuous-wave (cw) solution
\begin{equation}
u=u_0\exp\left(-\frac{2\ri}{q}|u_0|^2z\right), \quad n=\frac{1}{q}|u_0|^2,
\label{background}
\end{equation}
of Eqs.~(\ref{NLS1})-(\ref{NLS2}). Note that the constant amplitude
$u_0$ can be absorbed into $\rho$ in the Madelung transformation; nevertheless,
for convenience, we opt to use $u_0$ in Eq.~(\ref{Mand})
so that no extra free phase appears on the solutions that we present below,
thus making presentation more clear. To further elaborate on this point,
and also to underline the importance of this cw solution in our analysis
(because the cw defines the background on top of which our solutions will propagate),
%\tph{
%This cw solution is pivotal in our analysis as it defines
%the background on top of which our solutions will propagate. To make this more clear,
let us write the
solution of Eqs.~(\ref{NLS1})-(\ref{NLS2}) as
\[
u=U_0(z)\bar{u},\quad n=n_0\bar{n},
\]
where the background (cw) solution satisfies
\[
\mathrm{i}\frac{dU_0}{dz}=2n_0 U_0,\quad n_0=\frac{|U_0|}{q}.
\]
Obviously, the solution of the above equations is given in Eq.~(\ref{background}),
while $\bar{u}$ and $\bar{n}$
%Then the remaining terms follow
satisfy the system
\begin{eqnarray*}
\ri \bar{u}_{z} + \frac{1}{2} (\Delta \bar{u}- D\bar{u}_{tt})
-2\frac{|u_0|^2}{q}(1-\bar{n})\bar{u}=0,
\\
\frac{d}{q}\Delta\bar{n} - 2\bar{n} + 2|\bar{u}|^2 = 0.
\end{eqnarray*}
%
%which exhibits
This system possesses a %constant
cw solution of unit amplitude, as would be the case if $u_0$ was not used in the Madelung
transformation (i.e., in other words, $u_0$ will inevitably appear in the cw background solution).
%}

Below we seek nonlinear excitations (e.g., solitary waves) which propagate on top of this cw
background. It is, thus, relevant to investigate if this solution is subject to modulational
instability (MI): evidently, nonlinear excitations corresponding to an unstable background do not
have any physical purport.
The stability of the cw solution can be investigated upon employing Eqs.~(\ref{h1})-(\ref{h3})
as follows.
Let
%\tph{(cf. discussion on the background above)}
%
\[
\rho=1+\tilde{\rho},\quad \phi = - \frac{2}{q}|u_0|^2 z + \tilde{\phi}, \quad
n=\frac{1}{q}|u_0|^2+\tilde{n},
\]
where small perturbations $\tilde{\rho}$, $\tilde{\phi}$ and $\tilde{n}$ are
assumed to behave like
$\exp[\ri(k_z z + \boldsymbol{k}_{\perp} \cdot \boldsymbol{r}_{\perp} -\omega t)]$.
Here, we should recall that the evolution variable in our problem is the
propagation distance $z$ and, thus, the MI analysis is performed with respect
to this variable; as such, $k_z$ and its roots (real or imaginary) will determine
the stability of the cw solution.
To this end, substituting the above ansatz into Eqs.~(\ref{h1})-(\ref{h3}),
we find that %describe the evolution of
small-amplitude linear waves obey a dispersion relation of the following
form,
\begin{equation}
k_z^2=\frac{2|u_0|^2}{q}(\boldsymbol{k}_{\perp}^2-D\omega^2)
\left(1+\frac{d\boldsymbol{k}_{\perp}^2}{2q}\right)^{-1}
+ \frac{1}{4}(\boldsymbol{k}_{\perp}^2-D\omega^2)^2.
\label{dr}
\end{equation}
The results stemming from the above equation are as follows. First, Eq.~(\ref{dr}) shows
that the cw solution is always modulationally stable,
i.e., $k_z \in \mathbb{R}$~$\forall ~\boldsymbol{k}_{\perp},~\omega$, provided $D<0$.
Note, that in the $(1+1)$-dimensional case (corresponding to $k_{y}=0$ and $D=0$, or
$\boldsymbol{k}_{\perp}=0$ and $D=-1$), this result recovers the one of Ref.~\cite{krol1}.
Second, if $D=|D|>0$, the cw solution is unstable: in this case, perturbations
grow exponentially, with the instability growth rate given by ${\rm Im}(k_z)$.
%becomes . By adding further terms to this expansion will result in the
%collapse of the solution due to modulation instability.
%
Thus, hereafter, we focus %here
on this case, and assume that $D=-|D|$, corresponding to the anomalous GVD regime.
%%\tph{In the case where
%$D=|D|>0$, this solution is unstable. By adding further terms to this expansion will result in the
%collapse of the solution due to modulation instability. Also, and to avoid any confusion, note
%that, the MI analysis is performed with respect to the propagating variable $z$, as such $k_z$ and
%its roots (real or imaginary) will determine the systems stability.
%%}
It is, therefore, clear that the asymptotic analysis and results that we present in the
following Sections are only valid in this regime; in the opposite case, of $D=|D|>0$, since
the cw background is unstable, any small perturbations on top of it will result in
collapsing solutions.

Another physically relevant information stemming from Eq.~(\ref{dr}) is that, in the
long-wavelength and low-frequency limit (i.e., $|\boldsymbol{k}_{\perp}|,~\omega
\rightarrow
0$),
small-amplitude spatial or temporal waves propagate on top of the cw background with
``sound velocities'' $C^2$ or $V^2$, respectively, which are given by:
\begin{equation}
C^2 = \frac{2|u_0|^2}{q}, \qquad V^2=C^2 |D|.
%C = \pm |u_0|\sqrt{\frac{2}{q}}, \qquad V=C\sqrt{|D|} = \pm |u_0|\sqrt{\frac{2|D|}{q}}.
\label{cs}
\end{equation}
These characteristic velocities can also be determined, in a self-consistent
manner, in the framework of the reductive perturbation method (see, e.g., Ref.~\cite{djfbam}).
%and Appendix A).
It is also noted that in the unstable case of $D=|D|>0$, velocity $V$
becomes imaginary, a fact that also indicates that perturbations of the cw solution
grow exponentially in the propagation distance $z$.

\section{Effective nonlinear evolution equations}

\subsection{The Boussinesq equation}

Observing that the dispersion relation (\ref{dr}) resembles the one of a Boussinesq
equation \cite{MJA1,Johnson}, we now derive from Eqs.~(\ref{h1})-(\ref{h3}) a
Boussinesq equation, for either Cartesian or cylindrical geometry. We thus seek solutions
of Eqs.~(\ref{h1})-(\ref{h3}) in the form of the following asymptotic expansions:
\begin{equation}
\hspace*{-2.5cm}\phi=- \frac{2}{q}|u_0|^2 z +\vareps^{1/2}\Phi, \quad
%\rho=1+\sum_{j=1}^{\infty}\vareps^j \rho_j,
\rho=1+\vareps \rho_1 + \vareps^2 \rho_2 + \cdots, \quad
n= \frac{1}{q}|u_0|^2+ \vareps n_1 + \vareps^2 n_2  + \cdots,
%\sum_{j=1}^{\infty}\vareps^j n_j,
\label{expansions}
\end{equation}
where $\vareps$ is a formal small parameter ($0<\vareps \ll 1$), while
unknown real functions $\Phi$, $\rho_j$ and $n_j$ are assumed to
depend on ``slow variables''.
In particular, for the Cartesian geometry, $\Phi$, $\rho_j$ and $n_j$ are taken to
depend on $\{Z,~X,~Y,~T\}$, while, for the cylindrical geometry, on $\{Z,~R,~\theta,~T\}$
(the angular coordinate $\theta$ is assumed to remain unchanged);
these slow variables are defined as:
\begin{equation}
Z=\vareps^{1/2}z, \quad X=\vareps^{1/2}x, \quad
Y= \vareps^{1/2}y, \quad R=\vareps^{1/2}r, \quad T=\vareps^{1/2}t.
\label{slow}
\end{equation}
Substituting the expansions (\ref{expansions}) into Eqs.~(\ref{h1})-(\ref{h3}), and
using the variables in Eq.~(\ref{slow}), we obtain the following results.
First, Eq.~(\ref{h1}) reads:
\begin{eqnarray}
\hspace*{-2cm}\Phi_Z+2n_1+\vareps\left\{\frac{1}{2}\left[(\tilde{\boldsymbol{\nabla}}\Phi)^2+|D|\Phi_T^2\right]
-\frac{1}{4}\left(\tilde{\Delta}\rho_1+|D|\rho_{1TT}\right) +2n_2 \right\}
= \mathcal{O}(\vareps^2),
\label{Phi}
\end{eqnarray}
where
\[
\tilde{\Delta} = \partial_X^2 + \partial_Y^2, \qquad
\tilde{\boldsymbol{\nabla}}=(\partial_X,~\partial_Y)
\]
for the Cartesian case, while
\[
\tilde{\Delta} =\frac{1}{R}\partial_R(R\partial_R)+\frac{1}{R^2}\partial_{\theta}^2,
\qquad
\tilde{\boldsymbol{\nabla}} = \left(\partial_R,~\frac{1}{R}\partial_{\theta}\right)
\]
for the cylindrical case.
Second, Eq.~(\ref{h2}) leads, at orders $\mathcal{O}(\vareps^{3/2})$ and
$\mathcal{O}(\vareps^{5/2})$, to the following equations, respectively:
\begin{eqnarray}
\rho_{1Z}+\tilde{\Delta}\Phi+|D|\Phi_{TT}=0,
\label{nn1}
\\
\rho_{2Z} + \tilde{\boldsymbol{\nabla}} \cdot (\rho_1 \tilde{\boldsymbol{\nabla}}\Phi)
+|D| (\rho_1 \Phi_T)_T=0.
\label{nn2}
\end{eqnarray}
Finally, Eq.~(\ref{h3}), at orders $\mathcal{O}(\vareps)$ and
$\mathcal{O}(\vareps^{2})$, lead, respectively, to the equations:
\begin{eqnarray}
-2qn_1+2|u_0|^2\rho_1=0, \label{n1} \\
d \tilde{\Delta}n_1-2qn_2 +2|u_0|^2\rho_2=0.
\label{n2}
\end{eqnarray}
The leading-order part of Eq.~(\ref{Phi}), together with Eq.~(\ref{n1}),
provides the following connection between functions $\Phi$, $n_1$ and $\rho_1$:
\begin{equation}
\Phi_Z = -2n_1 = -C^2 \rho_1.
\label{Phir1}
\end{equation}
Furthermore, from the system of Eqs.~(\ref{Phi})-(\ref{n2}), it is possible to
eliminate functions $\rho_{1,2}$ and $n_{1,2}$, and derive the following equation for $\Phi$:
\begin{eqnarray}
&&\Phi_{ZZ}-C^2\left(\tilde{\Delta}\Phi + |D|\Phi_{TT}\right)
+\vareps\left\{\frac{1}{4C^2}\left(\alpha \tilde{\Delta}\Phi +|D| \Phi_{TT} \right)_{ZZ}
\right. \nonumber \\
&&\left.
+ \left[(\tilde{\boldsymbol{\nabla}}\Phi)^2+|D|\Phi_T^2\right]_{Z} %\right. \nonumber \\
+\Phi_Z \left(\tilde{\Delta}\Phi+|D|\Phi_{TT}\right)
\right\} = \mathcal{O}(\vareps^2),
\label{Bou}
\end{eqnarray}
where the parameter $\alpha$ is given by:
\begin{equation}
\alpha = 1-\frac{4d|u_0|^2}{q^2}.
\label{alpha}
\end{equation}

It is clear that, to leading-order, Eq.~(\ref{Bou}) is a linear wave equation
indicating that the velocities of spatial or temporal waves are indeed those
given in Eq.~(\ref{cs}). In addition, at order $\mathcal{O}(\vareps)$, Eq.~(\ref{Bou})
incorporates fourth-order dispersion terms and quadratic nonlinear terms.
Obviously, Eq.~(\ref{Bou}) is a Boussinesq-type equation, either in Cartesian
or cylindrical coordinates, as per the discussion above. The Boussinesq equation
has been originally proposed for studies of waves in shallow water
\cite{MJA1,Johnson,Karpman}, but later it was also used in different contexts,
ranging from ion-acoustic waves in plasmas \cite{Infeld,Karpman}
to mechanical lattices and electrical transmission lines \cite{Rem}.

A Boussinesq equation, similar to that in Eq.~(\ref{Bou}), was derived from
a $(2+1)$-dimensional NLS equation, in Cartesian coordinates, with a local defocusing
nonlinearity \cite{peli1}; earlier analysis of this Boussinesq model \cite{peli2} was
used in \cite{peli1} to investigate self-focusing and transverse instability
of plane dark solitons (see also \cite{pelirev} for a review and references therein).
In fact, the Cartesian version of the Boussinesq model~(\ref{Bou}) is reduced to
the one derived in \cite{peli1} in the limit of $d \rightarrow 0$, i.e., in the
local nonlinearity case.

\subsection{Kadomtsev-Petviashvilli-type equations}

We now proceed to derive the far-field equations stemming from the Boussinesq model
(\ref{Bou}), in the framework of multiscale asymptotic
expansions. As is known, the far-field of the Boussinesq equation
in $(1+1)$-dimensions is a pair of two KdV equations \cite{MJA1},
while in $(2+1)$-dimensions, it is a pair of KP equations
\cite{peli1}, for right- and left-going waves. Below we show that Eq.~(\ref{Bou})
gives rise to $(3+1)$-dimensional KP-type models for such waves. In addition,
we will distinguish cases corresponding to two different types of solitary waves
that may be supported  either in Cartesian or cylindrical geometry:
one, is oblique tube-shaped, oriented under a uniquely determined angle to
the propagation axis, i.e., a {\it spatial solitary wave}; the other, is
a constant-shape localized perturbation propagating along the $z$-axis,
i.e., a {\it temporal solitary wave}.

\subsubsection{Spatial solitary waves}

First, we consider spatial solitary waves, which may have either the form
of stripes propagating on the $XZ$ plane (Cartesian geometry), or exhibit
an annular shape, with the ring radius varying with the propagation distance
(cylindrical geometry). We thus introduce the variables:
\[
\chi = X-CZ, \quad \tilde{\chi}=X+CZ, \quad \mathcal{Z}=\vareps Z,
\quad \mathcal{Y}=\vareps^{1/2}Y, \quad \mathcal{T}=\vareps^{1/2}T,
\]
and
\[
\varrho = R-CZ, \quad \tilde{\varrho}= R+CZ, \quad \mathcal{Z}=\vareps Z,
\quad {\it \Theta}=\vareps^{-1/2}\theta, \quad \mathcal{T}=\vareps^{1/2}T,
\]
for the two geometries, respectively.
We also look for solutions of Eq.~(\ref{Bou}) in the form
of the asymptotic expansion:
\begin{equation}
%\Phi = \sum_{j=0}^{\infty} \vareps^j \Phi_j.
\Phi = \Phi_0 + \vareps \Phi_1 + \cdots.
\label{expphi}
\end{equation}
Substituting Eq.~(\ref{expphi}) into Eq.~(\ref{Bou}),
we obtain the following results. At leading-order,
$\mathcal{O}(1)$:
\begin{equation}
4C^2\Phi_{0\chi\tilde{\chi}}=0, \quad
4C^2\Phi_{0\varrho\tilde{\varrho}}=0,
\label{losp}
\end{equation}
for the Cartesian and cylindrical geometry, respectively. The
above equations imply that, in each case, $\Phi_0$ can be expressed as a superposition
of a right-going wave, $\Phi_0^{(R)}$, depending on $\chi$ or $\varrho$,
and a left-going one, $\Phi_0^{(L)}$, depending on $\tilde{\chi}$ or $\tilde{\varrho}$,
namely:
\begin{equation}
\Phi_{0}=\Phi_0^{(R)}+\Phi_0^{(L)}.
\label{rlcarsp}
\end{equation}
Second, at order $\mathcal{O}(\vareps)$:
\begin{eqnarray}
&&4C^2\Phi_{1\chi\tilde{\chi}} = -C\left(\Phi_{0\chi\chi}^{(R)}\Phi_{0\tilde{\chi}}^{(L)}
-\Phi_{0\chi}^{(R)}\Phi_{0\tilde{\chi}\tilde{\chi}}^{(L)} \right) \nonumber \\
&&+\left[
\left(-2C\Phi_{0\mathcal{Z}}^{(R)} +\frac{\alpha}{4}\Phi_{0\chi\chi\chi}^{(R)}
-\frac{3C}{2}\Phi_{0\chi}^{(R)2}\right)_{\chi}
-C^2\left(\Phi_{0\mathcal{Y}\mathcal{Y}}^{(R)}
+|D|\Phi_{0\mathcal{T}\mathcal{T}}^{(R)}\right)
\right]
\nonumber \\
&&+ \left[\left(2C\Phi_{0\mathcal{Z}}^{(L)}
+\frac{\alpha}{4}\Phi_{0\tilde{\chi}\tilde{\chi}\tilde{\chi}}^{(L)}
+\frac{3C}{2}\Phi_{0\tilde{\chi}}^{(L)2}
\right)_{\tilde{\chi}}
-C^2\left(\Phi_{0\mathcal{Y}\mathcal{Y}}^{(L)}
+|D|\Phi_{0\mathcal{T}\mathcal{T}}^{(L)}\right)
\right],
\label{phicarsp}
\end{eqnarray}
for the Cartesian geometry, and
\begin{eqnarray}
&&\hspace*{-1cm}4C^2\Phi_{1\varrho\tilde{\varrho}} =
-C \left(\Phi_{0\varrho\varrho}^{(R)}\Phi_{0\tilde{\varrho}}^{(L)}
-\Phi_{0\varrho}^{(R)}\Phi_{0\tilde{\varrho}\tilde{\varrho}}^{(L)}\right) \nonumber \\
&&\hspace*{-1cm}+\left[
\left(-2C\Phi_{0\mathcal{Z}}^{(R)}
+\frac{\alpha}{4}\Phi_{0\varrho\varrho\varrho}^{(R)}
-\frac{3C}{2}\Phi_{0\varrho}^{(R)2} - \frac{C}{\mathcal{Z}}\Phi_{0}^{(R)}\right)_{\varrho}
-\frac{1}{\mathcal{Z}^2} \Phi_{0{\it \Theta}{\it \Theta}}^{(R)}
-V^2 \Phi_{0\mathcal{T}\mathcal{T}}^{(R)}
\right]
\nonumber \\
&&\hspace*{-1cm}+\left[
\left(2C\Phi_{0\mathcal{Z}}^{(L)}
+\frac{\alpha}{4}\Phi_{0\tilde{\varrho}\tilde{\varrho}\tilde{\varrho}}^{(L)}
+\frac{3C}{2}\Phi_{0\tilde{\varrho}}^{(L)2}
-\frac{C}{\mathcal{Z}}\Phi_{0}^{(L)}\right)_{\tilde{\varrho}}
-\frac{1}{\mathcal{Z}^2} \Phi_{0{\it \Theta}{\it \Theta}}^{(R)}
-V^2 \Phi_{0\mathcal{T}\mathcal{T}}^{(L)}
\right].
\label{phicylspl}
\end{eqnarray}
for the cylindrical geometry.
Upon integrating Eq.~(\ref{phicarsp}) in $\chi$ or $\tilde{\chi}$
[Eq.~(\ref{phicylspl}) in $\varrho$ or $\tilde{\varrho}$],
%in the regular or in the $\sim$ variables,
it is obvious that the terms in square brackets in the right-hand side
%of these equations
are secular, because
are functions of $\chi$ or $\tilde{\chi}$ (of $\varrho$ or $\tilde{\varrho}$)
alone. Removal of these terms leads to two uncoupled nonlinear evolution
equations for $\Phi_0^{(R)}$ and $\Phi_0^{(L)}$. Furthermore, employing
Eq.~(\ref{Phir1}), it is straightforward to find that the amplitude
$\rho_1$ can also be decomposed to a left- and a right-going wave,
i.e., $\rho_1 = \rho_1^{(R)}+\rho_1^{(L)}$, with
\[
\hspace*{-1cm}\Phi_{0\chi}^{(R)}=C\rho_1^{(R)}, \quad \Phi_{0\tilde{\chi}}^{(L)}=-C\rho_1^{(L)},
\quad {\rm and} \quad
\Phi_{0\varrho}^{(R)}=C\rho_1^{(R)}, \quad \Phi_{0\tilde{\varrho}}^{(L)}=-C\rho_1^{(L)}.
\]
Then, using the above expressions, the equations for $\Phi_0^{(R)}$
and $\Phi_0^{(L)}$ yield, in each geometry, two uncoupled equations for $\rho_1^{(R)}$
and $\rho_1^{(L)}$. In Cartesian geometry, these equations are:
\begin{eqnarray}
\left(\rho_{1\mathcal{Z}}^{(R)} -\frac{\alpha}{8C}\rho_{1\chi\chi\chi}^{(R)}
+\frac{3C}{2}\rho_{1}^{(R)} \rho_{1\chi}^{(R)}\right)_{\chi}
+\frac{C}{2}\left(\rho_{1\mathcal{Y}\mathcal{Y}}^{(R)}
+|D|\rho_{1\mathcal{T}\mathcal{T}}^{(R)}\right)=0,
\label{KPcarspR} \\
\left(\rho_{1\mathcal{Z}}^{(L)}
+\frac{\alpha}{8C}\rho_{1\tilde{\chi}\tilde{\chi}\tilde{\chi}}^{(L)}
-\frac{3C}{2}\rho_{1}^{(L)} \rho_{1\tilde{\chi}}^{(L)}\right)_{\tilde{\chi}}
-\frac{C}{2}\left(\rho_{1\mathcal{Y}\mathcal{Y}}^{(L)}
+|D|\rho_{1\mathcal{T}\mathcal{T}}^{(L)}\right)=0.
\label{KPcarspL}
\end{eqnarray}
On the other hand, equations for $\rho_{1}^{(R,L)}$ in cylindrical geometry, are:
\begin{eqnarray}
\hspace*{-2cm}\left(\rho_{1\mathcal{Z}}^{(R)}
-\frac{\alpha}{8C}\rho_{1\varrho\varrho\varrho}^{(R)}
+\frac{3C}{2}\rho_{1}^{(R)}\rho_{1\varrho}^{(R)} +
\frac{1}{2\mathcal{Z}}\rho_{1}^{(R)}\right)_{\varrho}
+\frac{1}{2C}\left( \frac{1}{\mathcal{Z}^2} \rho_{1{\it \Theta}{\it \Theta}}^{(R)}
+V^2 \rho_{1\mathcal{T}\mathcal{T}}^{(R)}\right) =0,
\label{rcylspr} \\
\hspace*{-2cm}\left(\rho_{1\mathcal{Z}}^{(L)}
+\frac{\alpha}{8C}\rho_{1\tilde{\varrho}\tilde{\varrho}\tilde{\varrho}}^{(L)}
-\frac{3C}{2}\rho_{1}^{(L)}\rho_{1\tilde{\varrho}}^{(L)}
-\frac{1}{2\mathcal{Z}}\rho_{1}^{(L)}\right)_{\tilde{\varrho}}
-\frac{1}{2C}\left( \frac{1}{\mathcal{Z}^2} \rho_{1{\it \Theta}{\it \Theta}}^{(L)}
+V^2 \rho_{1\mathcal{T}\mathcal{T}}^{(L)}\right) =0.
\label{rcylspl}
\end{eqnarray}

\subsubsection{Temporal solitary waves}

We now proceed with the case of temporal solitary waves. First, introduce the variables:
\[
\tau = T-VZ, \quad \tilde{\tau}=T+VZ, \quad \mathcal{Z}=\vareps Z,
\]
as well as
\[
\mathcal{X}=\vareps^{1/2}X, \quad \mathcal{Y}=\vareps^{1/2}Y,
\quad {\rm and} \quad
\mathcal{R}=\vareps^{1/2}R, \quad {\it \Theta}=\vareps^{-1/2}\theta,
\]
for the Cartesian and cylindrical geometry, respectively.
Then, in each case, utilizing the above variables and the asymptotic
expansion~(\ref{expphi}),
we obtain from Eq.~(\ref{Bou}) the leading-order equation:
\[
4V^2\Phi_{0\tau\tilde{\tau}}=0,
\]
which yields again Eq.~(\ref{rlcarsp}). Furthermore,
working as in the previous case, we obtain at order $\mathcal{O}(\vareps)$:
\begin{eqnarray}
&&4V^2\Phi_{1\tau\tilde{\tau}} = -V|D|
\left(\Phi_{0\tau\tau}^{(R)}\Phi_{0\tilde{\tau}}^{(L)}
-\Phi_{0\tau}^{(R)}\Phi_{0\tilde{\tau}\tilde{\tau}}^{(L)}\right) \nonumber \\
&&+\left[
\left(-2V\Phi_{0\mathcal{Z}}^{(R)} +\frac{D^2}{4}\Phi_{0\tau\tau\tau}^{(R)}
-\frac{3|D|V}{2}\Phi_{0\tau}^{(R)2}\right)_{\tau}
-C^2 \hat{\Delta}\Phi_{0}^{(R)} \right],
\nonumber \\
&&+\left[
\left(2V\Phi_{0\mathcal{Z}}^{(L)}
+\frac{D^2}{4}\Phi_{0\tilde{\tau}\tilde{\tau}\tilde{\tau}}^{(L)}
+\frac{3|D|V}{2}\Phi_{0\tilde{\tau}}^{(L)2}
\right)_{\tilde{\tau}}
-C^2 \hat{\Delta}\Phi_{0}^{(L)}
\right],
\label{phicarteml}
\end{eqnarray}
where
\[
\hat{\Delta} =
\partial_{\mathcal{X}}^2+\partial_{\mathcal{Y}}^2, \quad
\hat{\Delta}=
\frac{1}{\mathcal{R}}\partial_\mathcal{R}(\mathcal{R}\partial_\mathcal{R})
+\frac{1}{\mathcal{R}^2}\partial_{{\it \Theta}}^2,
\]
for the two geometries, respectively.
Then, employing Eq.~(\ref{Phir1}), the amplitude
$\rho_1$ is again expressed as $\rho_1 = \rho_1^{(R)}+\rho_1^{(L)}$, with
\begin{equation}
\Phi_{0\tau}^{(R)}=\frac{C^2}{V}\rho_1^{(R)}, \quad
\Phi_{0\tilde{\tau}}^{(L)}=-\frac{C^2}{V}\rho_1^{(L)}.
\label{phirte}
\end{equation}
Using Eqs.~(\ref{phirte}), we obtain from Eq.~(\ref{phicarteml})
the following equations for $\rho_1^{(R,L)}$:
\begin{eqnarray}
\left(\rho_{1\mathcal{Z}}^{(R)} -\frac{D^2}{8V}\rho_{1\tau\tau\tau}^{(R)}
+\frac{3V}{2}\rho_{1}^{(R)} \rho_{1\tau}^{(R)}\right)_{\tau}
+\frac{V}{2|D|}\hat{\Delta}\rho_{1}^{(R)}=0,
\label{KPcarteR} \\
\left(\rho_{1\mathcal{Z}}^{(L)}
+\frac{D^2}{8V}\rho_{1\tilde{\tau}\tilde{\tau}\tilde{\tau}}^{(L)}
-\frac{3V}{2}\rho_{1}^{(L)} \rho_{1\tilde{\tau}}^{(L)}\right)_{\tilde{\tau}}
-\frac{V}{2|D|}\hat{\Delta}\rho_{1}^{(L)}=0.
\label{KPcarteL}
\end{eqnarray}

We conclude this section with the observation that all equations that were derived
for $\rho_{1}^{(R,L)}$ are of the KP type, in both geometries.
Below we elaborate more on these effective models, and
focus on limiting cases corresponding to their lower-dimensional versions.
For simplicity, we only consider the right-going
waves, $\rho_{1\mathcal{Z}}^{(R)}$, since
$\rho_{1}^{(L)}(\mathcal{Z})=\rho_{1}^{(R)}(-\mathcal{Z})$.
In addition, we will present examples of solitary wave solutions of
Eqs.~(\ref{NLS1})-(\ref{NLS2}) arising from these KP models.

\section{Versions of the KP equations and solitary waves}

\subsection{Classification of the effective KP models}

First of all, it is convenient to further normalize the effective KP equations
derived in the previous section in order to express them in their
``standard'' form \cite{BlackBook,Infeld}.

Consider, first, equations for spatial solitary waves, and introduce
the transformations:
\begin{equation}
\mathcal{Z} \rightarrow -\frac{\alpha}{8C}\mathcal{Z},
\quad
\mathcal{T} \rightarrow \sqrt{\frac{3|\alpha|}{4V^2}}\mathcal{T},
\label{toKP1}
\end{equation}
as well as
\[
\hspace*{-1cm}\mathcal{Y} \rightarrow \sqrt{\frac{3|\alpha|}{4C^2}}\mathcal{Y},
\quad
\rho_1^{(R)} = -\frac{\alpha}{2C^2}U
\quad {\rm and} \quad
{\it \Theta} \rightarrow \sqrt{\frac{3|\alpha|}{4}}{\it \Theta},
\quad
\rho_1^{(R)} = -\frac{\alpha}{2C^2}W,
\]
for the Cartesian and cylindrical geometry, respectively. This way,
Eq.~(\ref{KPcarspR}) is expressed as:
\begin{eqnarray}
\left(U_{\mathcal{Z}}+6UU_{\chi}+U_{\chi\chi\chi} \right)_{\chi}
+3\sigma^2 \left(U_{\mathcal{Y}\mathcal{Y}}+U_{\mathcal{T}\mathcal{T}}\right)=0,
\label{usKP}
\end{eqnarray}
while Eq.~(\ref{rcylspr}) reads:
\begin{eqnarray}
\left(W_{\mathcal{Z}}+6WW_{\varrho}+W_{\varrho\varrho\varrho}
+\frac{1}{2\mathcal{Z}}W\right)_{\varrho}
+3\sigma^2 \left(\frac{1}{\mathcal{Z}^2}W_{{\it \Theta}{\it
\Theta}}+W_{\mathcal{T}\mathcal{T}}\right)=0.
\label{usJ}
\end{eqnarray}
In the above equations, parameter $\sigma^2$ is given by
\[
\sigma^2=-\sgn{\alpha},
\]
and it is reminded that $\alpha$ is given by Eq.~(\ref{alpha}).

The $(1+1)$-dimensional versions of Eqs.~(\ref{usKP}) and (\ref{usJ}),
i.e., the ones referring to the $\mathcal{Z}\chi$ and $\mathcal{Z}\varrho$ plane,
have respectively the form of a KdV and a cylindrical KdV (cKdV) equation.
Both models are completely integrable by means of the IST \cite{BlackBook},
and find numerous applications in a variety of physical contexts
\cite{MJA1,Rem,Johnson,Infeld}.
The KdV and cKdV equations have been derived by means of multiscale expansion methods
from local NLS models, with the aim to describe
shallow planar dark solitons in Bose gases \cite{tsu} and
ring dark solitons in nonlinear optical media
\cite{ofyrds} (see also reviews \cite{ofy,djf} and references therein).
More recently, a KdV equation was derived from the $(1+1)$-dimensional version
of Eqs.~(\ref{NLS1})-(\ref{NLS2}) for $D=0$, and used to describe small-amplitude
nematicons \cite{tph}; in fact, the KdV model derived in \cite{tph} is identical with
the $(1+1)$-dimensional version of Eq.~(\ref{KPcarspR}) [or (\ref{usKP})].

Furthermore, there are two distinct $(2+1)$-dimensional versions of
Eq.~(\ref{usKP}): a spatial one, in the
$\mathcal{Z}\chi\mathcal{Y}$ space, and a spatio-temporal one,
in the $\mathcal{Z}\chi\mathcal{T}$ space. These effective models
can be used to describe either spatial optical solitons in nematic liquid
crystals \cite{ass1}, or dispersion-induced
dynamics of spatial solitons in thermal media \cite{krol1}. Both these
$(2+1)$-dimensional equations, are completely integrable by means of the IST
\cite{BlackBook}.

Importantly, the $(2+1)$-dimensional versions, as well as the complete Eq.~(\ref{usKP}),
include both versions of the KP equation, KP-I and KP-II \cite{BlackBook}.
Indeed, for $\sigma=1$, i.e., $\alpha<0\Rightarrow d>(q/2|u_0|)^2$, Eq.~(\ref{usKP})
is a KP-II equation; on the other hand, for $\sigma=\ri$, i.e.,
$\alpha>0\Rightarrow d<(q/2|u_0|)^2$, Eq.~(\ref{usKP}) is a KP-I equation.
Recalling that $d$ is the degree of nonlocality of the system at
hand (for $d \rightarrow 0$ nonlocal NLS Eqs.~(\ref{NLS1})-(\ref{NLS2}) become local),
it is evident that relatively weak (strong) nonlocality, as defined by the above
regimes of $d$, corresponds to a KP-I (KP-II) model.
This fact has also important implications on the type and the stability of
low-dimensional solitary waves that can be supported in the system (see below).

Similarly, we observe that there are two distinct $(2+1)$-dimensional versions of
Eq.~(\ref{usJ}): a spatial one, in the
$\mathcal{Z}\varrho{\it \Theta}$ space, and a spatio-temporal one,
in the $\mathcal{Z}\varrho\mathcal{T}$ space, which find applications in
the contexts discussed above in the Cartesian case. The spatial version of
Eq.~(\ref{usJ}) is a cylindrical KP (cKP) equation, which is
also known as the Johnson's equation \cite{J1}, and describes
nearly-concentric solitons in an ideal, inviscid fluid \cite{Johnson}. This model
is, also, completely integrable by means of the IST \cite{oevel}. On the other hand,
in the $\mathcal{Z}\varrho\mathcal{T}$ space, Eq.~(\ref{rcylspr})
reduces to the so-called CI equation, which describes weak cylindrical
ion-acoustic solitons in plasmas \cite{Infeld}. Unlike the Johnson's equation,
the CI equation is not considered to be integrable, as it fails to pass the Painlev\'e test
\cite{BlackBook}.

It is interesting to point out that there exist transformations mapping solutions
of the KP and cKP equations \cite{Johnson}. Indeed, the map:
\[
U(\mathcal{Z},~\chi,~\mathcal{Y})
%~\mathcal{T})
\rightarrow
W(\mathcal{Z},~\varrho,~{\it \Theta})
%,~\mathcal{T})
:=U\left(\mathcal{Z},~\varrho-\frac{\mathcal{Z}{\it \Theta}^2}{12\sigma^2},
~\mathcal{Z}{\it \Theta}
%,~\mathcal{T}
\right),
\]
transforms any solution of the KP equation~(\ref{usKP}) into a
solution of the cKP equation~(\ref{usJ}); conversely, the map:
\[
W(\mathcal{Z},~\chi,~{\it \Theta})
%,~\mathcal{T})
\rightarrow
U(\mathcal{Z},~\chi,~\mathcal{Y})
%,~\mathcal{T})
:= W\left(\mathcal{Z},~\chi+\frac{\mathcal{Y}^2}{12\sigma^2\mathcal{Z}},
~\frac{\mathcal{Y}}{\mathcal{Z}}
%,~\mathcal{T}
\right),
\]
transforms any solution of the cKP equation~(\ref{usJ}) into a
solution of the KP equation~(\ref{usKP}). Here, we should also note that
the spatial $(2+1)$-dimensional versions of KP Eq.~(\ref{usKP}) and cKP Eq.~(\ref{usJ})
are also connected with another relevant model, the elliptic cKP (ecKP),
that was recently presented and studied in \cite{mat}. In this work, it was
shown that the ecKP model describes surface gravity waves of nearly elliptic
fronts,
and it is completely integrable. Based on the similarities
of the hydrodynamic form (\ref{h1})-(\ref{h3}) of the nonlocal NLS
Eqs.~(\ref{NLS1})-(\ref{NLS2}) to the problem formulation of \cite{mat},
we conjecture that, adopting an elliptic cylindrical coordinate system and
following the lines of the analysis presented here, one
could derive a $(3+1)$-dimensional version of the ecKP equation. Nevertheless,
such a derivation is beyond the scope of the present work.

We now turn our attention to the KP models that describe temporal waves.
As before, first we put Eqs.~(\ref{KPcarteR}) in the ``standard'' form. We thus
introduce the transformations:
\begin{equation}
\hspace*{-1cm}\mathcal{Z} \rightarrow -\frac{D^2}{8V}\mathcal{Z},
\quad
\{ \mathcal{X},~\mathcal{Y},~\mathcal{R} \} \rightarrow
\sqrt{\frac{3|D|^3}{4V^2}} \{ \mathcal{X},~\mathcal{Y},~\mathcal{R}\},
\quad
\rho_1^{(R)} = -\frac{D^2}{2V^2}Q,
\label{toKPt1}
\end{equation}
and obtain from Eqs.~(\ref{KPcarteR}) the models:
\begin{eqnarray}
\left(Q_{\mathcal{Z}} +6QQ_{\tau} +Q_{\tau\tau\tau} \right)_{\tau}
-3 \hat{\Delta}Q=0,
\label{KPte}
\end{eqnarray}
and it is reminded that the Laplacian $\hat{\Delta}$ refers to either the Cartesian or
the cylindrical geometry. Obviously, the $(1+1)$-dimensional version of Eq.~(\ref{KPte})
is the KdV equation. On the other hand, it is observed that, unlike the case of spatial
solitary waves, the Cartesian version of Eq.~(\ref{KPte}) is solely of the KP-I type;
in fact, in this case, transverse effects are not governed by the sign of parameter
$\alpha$.
The $(2+1)$-dimensional version of Eq.~(\ref{KPte}) is completely integrable
by means of the IST \cite{BlackBook}. Finally, the cylindrical version of Eq.~(\ref{KPte})
is known as the CII equation, and describes cylindrical ion-acoustic solitons in plasmas
\cite{Infeld}.

\subsection{Solitary wave solutions}

The asymptotic reduction of the nonlocal NLS equations to the effective equations
above, allows for the derivation of approximate solutions of
Eqs.~(\ref{NLS1})-(\ref{NLS2}),
valid up to -- and including -- order $\mathcal{O}(\vareps)$.
Of particular interest are solitary wave solutions,
which can be constructed from solutions of Eqs.~(\ref{usKP}), (\ref{usJ}) and (\ref{KPte}).
These asymptotic reductions provide information on the type of the
solitary wave, as well as on the stability of lower-dimensional solutions in
higher-dimensional
settings. Below, we showcase some characteristic examples along those lines.

Let us first consider the case of spatial solitary waves. The $(1+1)$-dimensional version
of Eq.~(\ref{usKP}) is a KdV equation which possesses the commonly known soliton solution:
\begin{equation}
U=2\kappa^2 {\rm sech}^2[\kappa(\chi-4\kappa^2\mathcal{Z}-\chi_0)],
\label{carsol}
\end{equation}
where $\kappa$ and $\chi_0$ are constants. Using this solution, and reverting
transformations
for the independent variables and fields, we find the following approximate solution
to Eqs.~(\ref{NLS1})-(\ref{NLS2}):
\begin{eqnarray}
u &\approx& u_0 \left[1-\frac{\vareps \kappa^2}{C^2} \alpha {\rm sech}^2(\xi)\right]
\exp\left[-\frac{2\ri}{q}|u_0|^2z -\ri  \frac{\vareps^{1/2}\kappa }{C}
\alpha \tanh(\xi) \right],
\label{uspc} \\
n &\approx& \frac{1}{q} + \frac{1}{2}\vareps \kappa^2 \alpha {\rm sech}^2(\xi),
\label{nspc} \\
\xi &\equiv& \vareps^{1/2}\kappa (x-\upsilon_{\rm s}z-x_0),
\quad
\upsilon_{\rm s} \equiv C\left(1-\frac{1}{2}\frac{\vareps \eta^2}{C^2}|\alpha| \right).
\label{orsca}
\end{eqnarray}
The solution for $u$ has the form of either a density dip (for $\alpha>0$) or a
density hump (for $\alpha<0$) on top of the cw background, with a $\tanh$-shaped phase jump
across the density minimum or maximum, respectively. It is thus either a dark
soliton (for $\alpha>0$) or an anti-dark soliton (for $\alpha<0$); note that the
soliton velocity $\upsilon_{\rm s}$ is slightly below the speed of sound, as is the case
of shallow dark solitons in local media \cite{ofy,djf}. Note that if $d \rightarrow 0$,
then $\alpha>0$, which means that in the case of the local system the soliton is always
dark. In other words, anti-dark solitons are only supported due to the presence of nonlocality,
in accordance with the analysis of \cite{tph}.

%The fact that Eq.~(\ref{usKP}) is either a KP-I (for $\alpha<0$) or a KP-II (for
%$\alpha>0$) equation, can be used to deduce stability of the approximate solitons in
%$(2+1)$-dimensions. Indeed, as is well known \cite{BlackBook}, line soliton solutions
%of KP-I are unstable, while those of KP-II are stable. This leads to the prediction that,
%in the context of the original problem, dark soliton stripes of the nonlocal problem
%will be unstable in the 2D setting, while anti-dark soliton stripes will be stable.
%Note that the instability in the context of the KP-I model was analyzed \cite{peli2}
%and connected to the context of self-focusing and transverse instability
%of plane dark solitons in media with local defocusing nonlinearity \cite{kuz2,peli1}
%(see also \cite{pelirev} for a review and references therein). It should also be mentioned
%that in the case of KP-I (for $\alpha<0$), there exist ``lump'' solitons which are stable
%in the 2D setting \cite{BlackBook}; these structures can be used to construct approximate
%solutions of the original problem which, in our case, will be 2D dark solitary waves,
%featuring an algebraic decay. These ``lump'' solitons, however (along with the planar
%ones discussed above), are unstable in the full $(3+1)$-dimensional setting \cite{kuz1}.

\begin{figure}[tbp]
\centering
%[height=5.5cm]
\includegraphics[scale=0.37]{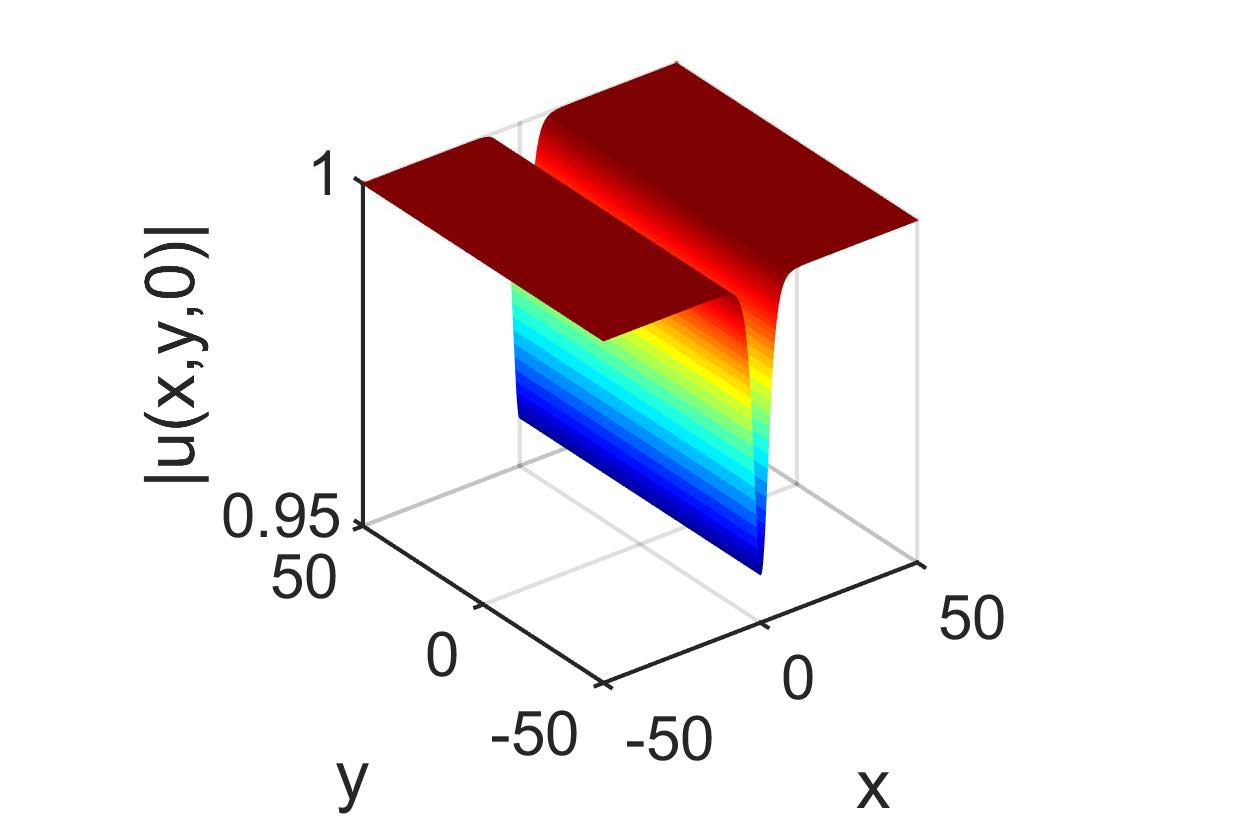}
%[height=5.5cm]
\includegraphics[scale=0.37]{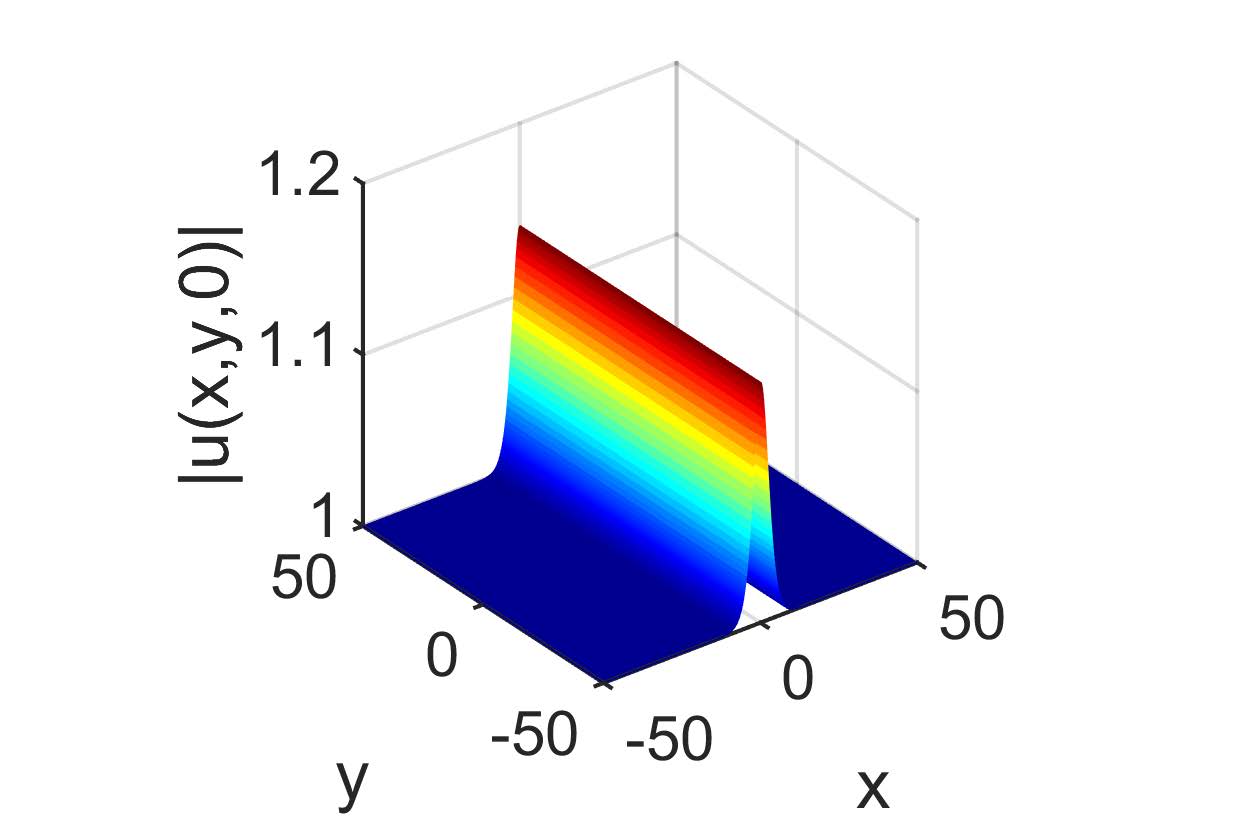}
\caption{(Color Online) Typical dark (left) and anti-dark soliton (right) profiles, at $z=0$,
in Cartesian geometry, for $q=1$ and $q=5$, respectively. All other parameter values are equal to unity.}
\label{stripes}
\end{figure}

\begin{figure}[ht]
\centering
\includegraphics[height=4cm]{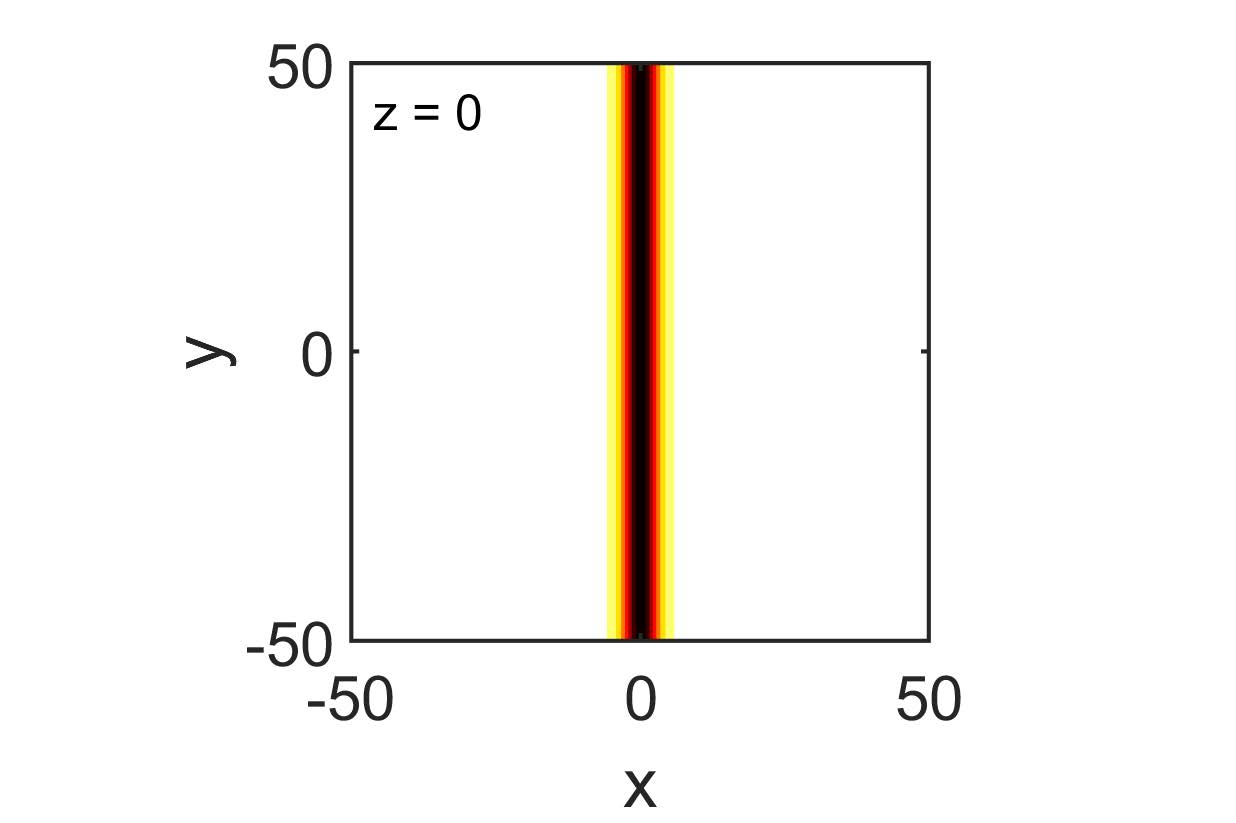}\hspace*{-1.2cm}
\includegraphics[height=4cm]{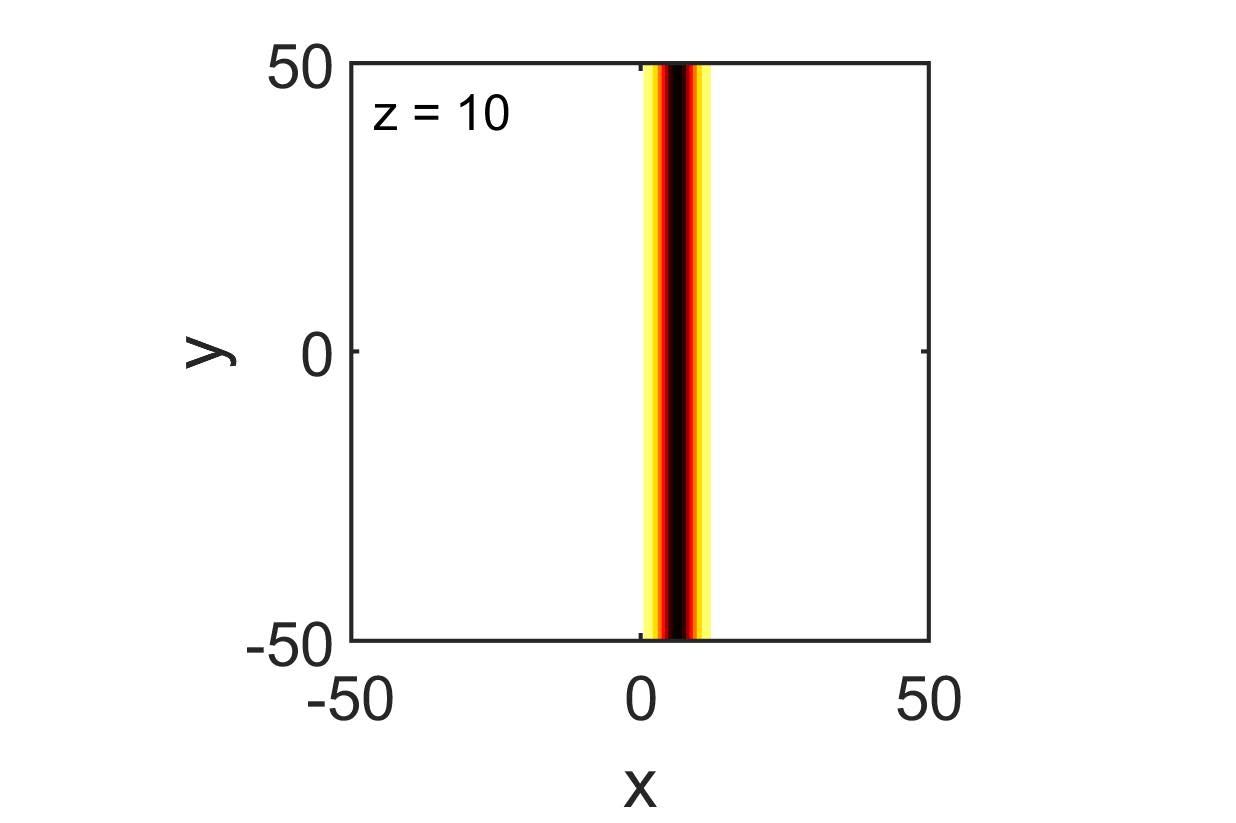}\hspace*{-1.2cm}
\includegraphics[height=4cm]{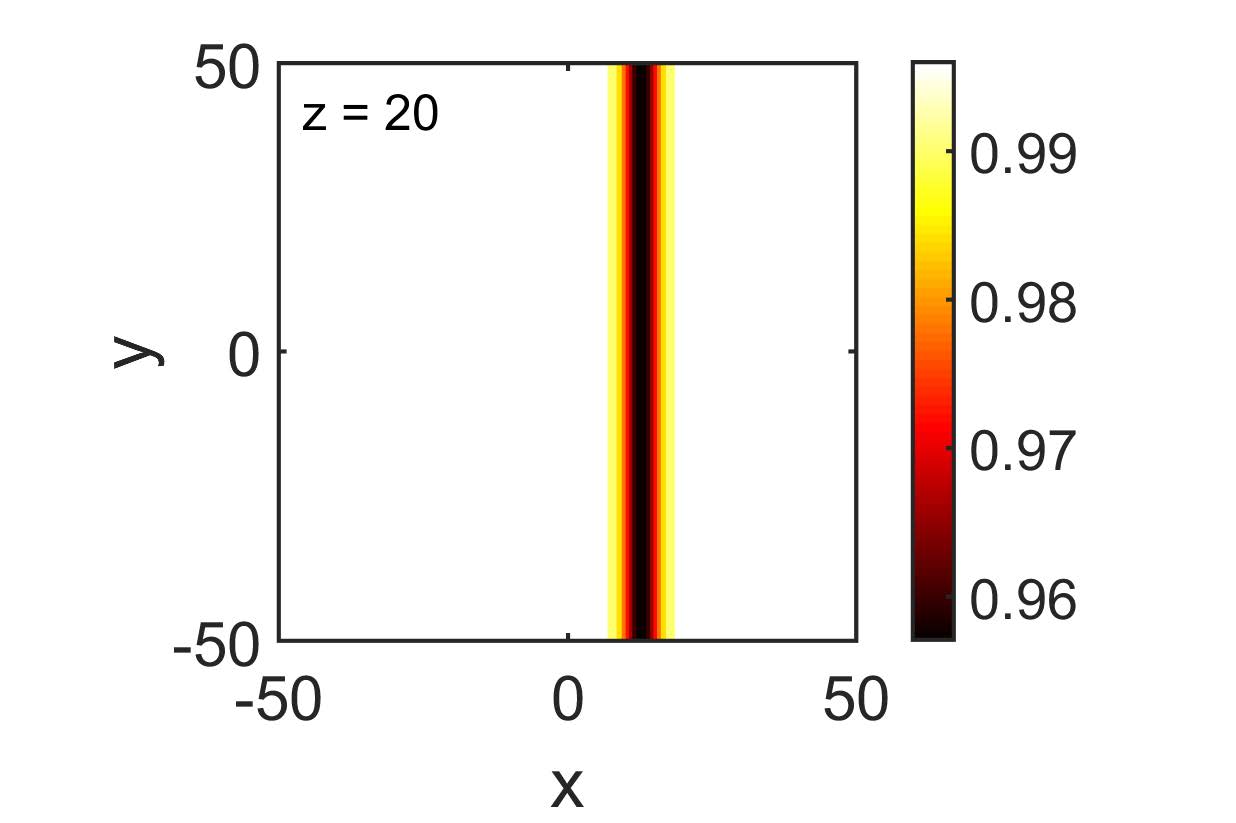}\\
\includegraphics[height=4cm]{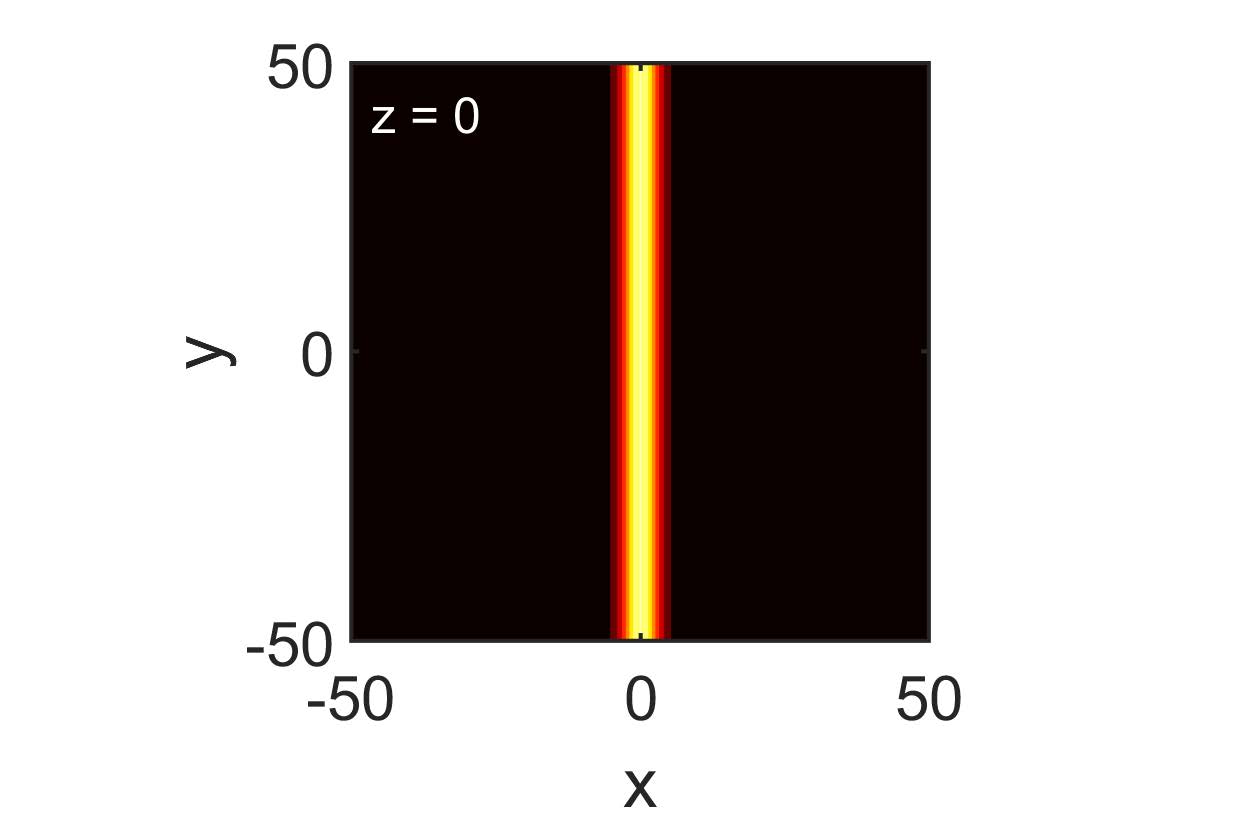}\hspace*{-1.2cm}
\includegraphics[height=4cm]{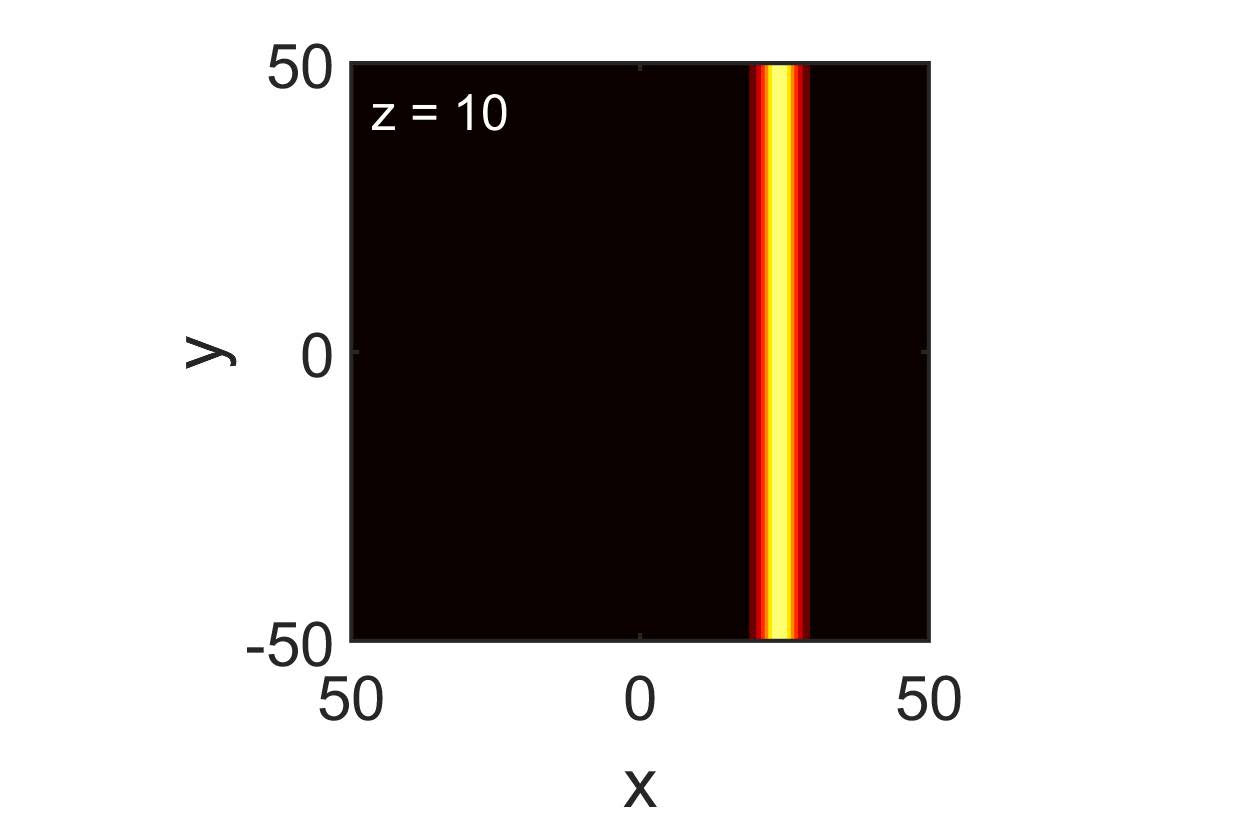}\hspace*{-1.2cm}
\includegraphics[height=4cm]{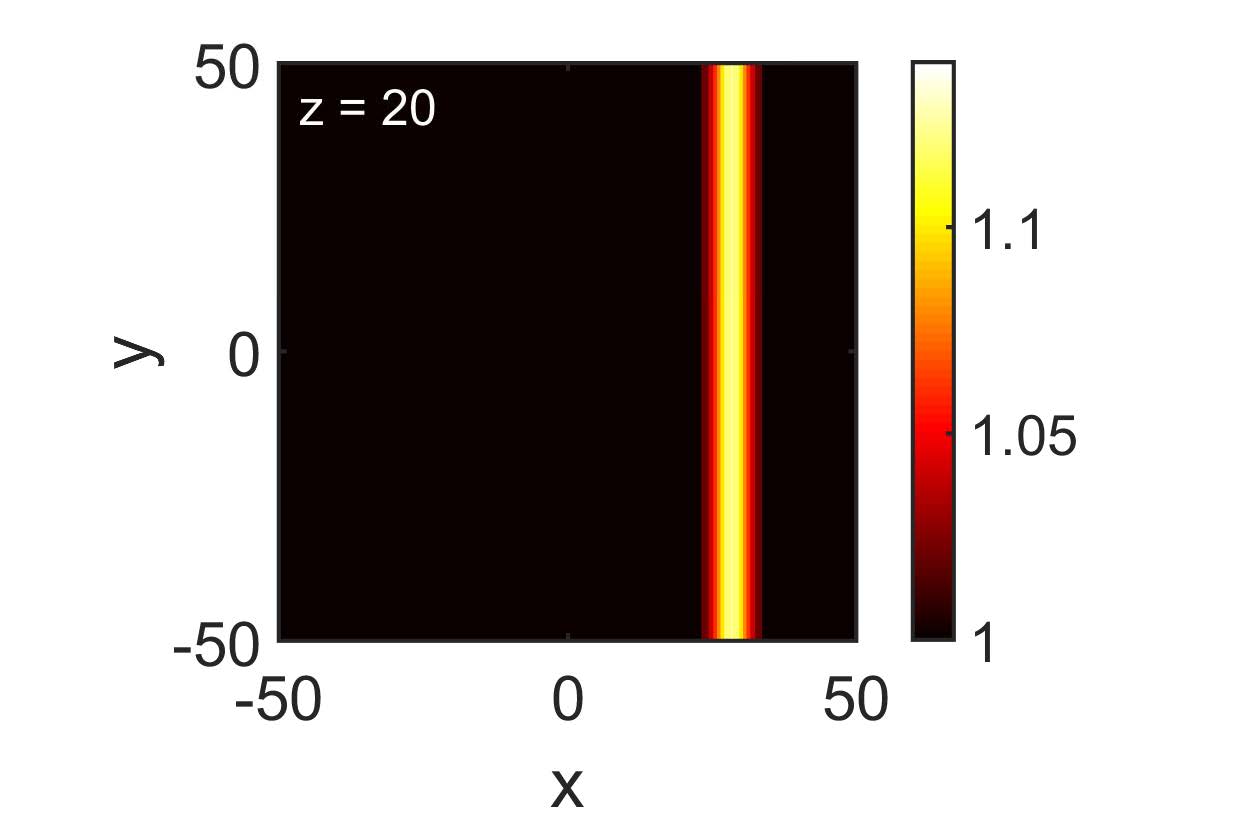}
\caption{(Color Online) Contour plots showing the evolution of the dark (top) and anti-dark
(bottom) solitons of Fig.~\ref{stripes}. Results have been obtained from direct
numerical integration of Eqs.~(\ref{NLS1})-(\ref{NLS2}).}
\label{stripes_evol}
\end{figure}

%\tph{
In Fig.~\ref{stripes}, we depict the soliton solutions in Cartesian geometry according to
Eq.~(\ref{carsol}). Here, solutions' profiles are plotted at $z=0$; all parameter values
%relative constants
are kept equal to unity, and we vary parameter $q$ so that to obtain a dark
%($q=1$)
and an anti-dark
%($q=5$)
soliton, for $q=1$ and $q=5$, respectively.
%can be obtained.
%}
Furthermore,
%to verify that these solutions maintain their stability and propagating characteristics,
we use these profiles as initial conditions, and perform a direct
numerical integration of Eqs.~(\ref{NLS1})-(\ref{NLS2}) to determine their evolution.
For the simulations, we used a high accuracy spectral integrator
in Cartesian coordinates.
%evolve these solutions according to Eqs.~(\ref{NLS1})-(\ref{NLS2}), cf
The results are shown in the contour plots of Fig.~\ref{stripes_evol}, where it is
verified that these solutions maintain their stability --
at least for relatively short propagation distances (see discussion below) --
and propagating characteristics.
Notice that, as expected from the analysis, the anti-dark soliton propagates at higher,
though constant, velocity from its dark soliton counterpart.

%\tph{
%To verify that these solutions maintain their stability and propagating characteristics, we
%evolve these solutions according to Eqs.~(\ref{NLS1})-(\ref{NLS2}), cf Fig. \ref{stripes_evol}.
%}

%\tph{ Notice that as expected from the analysis, the anti-dark soliton propagates at higher, though
%constant, velocity from its dark counterpart.}

The fact that Eq.~(\ref{usKP}) is either a KP-I (for $\alpha<0$) or a KP-II (for
$\alpha>0$) equation, can be used to deduce stability of the approximate solitons in
$(2+1)$-dimensions. Indeed, as is well known \cite{BlackBook}, line soliton solutions
of KP-I are unstable, while those of KP-II are stable. This leads to the prediction that,
in the context of the original problem, dark soliton stripes of the nonlocal problem
will be unstable in the 2D setting, while anti-dark soliton stripes will be stable.
Note that the instability in the context of the KP-I model was analyzed \cite{peli2}
and connected to the context of self-focusing and transverse instability
of plane dark solitons in media with local defocusing nonlinearity \cite{kuz2,peli1}
(see also \cite{pelirev} for a review and references therein). It should also be mentioned
that in the case of KP-I (for $\alpha<0$), there exist ``lump'' solitons which are stable
in the 2D setting \cite{BlackBook}; these structures can be used to construct approximate
solutions of the original problem which, in our case, will be 2D dark solitary waves,
featuring an algebraic decay. These ``lump'' solitons, however (along with the planar
ones discussed above), are unstable in the full $(3+1)$-dimensional setting \cite{kuz1}.

We now turn to the case of the cylindrical geometry, and consider the $(1+1)$-dimensional
version of Eq.~(\ref{usJ}), namely the cKdV equation. As mentioned above, this model is
completely integrable by means of the IST. The solitary wave solution, which is expressed
in terms of the Airy function \cite{hirota}, is composed of a primary wave and a shelf.
An asymptotic analysis \cite{jwm,ko} in the regime
$|\mathcal{Z}|\gg |\varrho|$ shows the following: to leading-order approximation,
the primary wave $W(\varrho,\mathcal{Z})$, that decays to zero at both upstream and
downstream infinity, has a form similar to that of Eq.~(\ref{carsol}), with the obvious
changes $\chi \rightarrow \varrho$ and $\chi_0 \rightarrow \varrho_0$, but with an
important difference:
$\kappa$ now becomes a slowly-varying function of $\mathcal{Z}$,
due to the presence of the term $W/(2\mathcal{Z})$. In fact, according to
the analysis of Refs.~\cite{ko,jwm}, and using the original coordinates, the following
result can be obtained,
\begin{equation}
\kappa^2=\kappa_0^2\left(\frac{z_0}{z} \right)^{2/3},
\label{kapa}
\end{equation}
where $\kappa_0^2$ is a constant setting the solitary wave amplitude at $z=z_0$.
Then, it is straightforward to express an approximate solution of
Eqs.~(\ref{NLS1})-(\ref{NLS2}), but now for the cylindrical geometry, and
for the primary solitary wave. This is of the form of Eqs.~(\ref{uspc})-(\ref{orsca}),
but with the solitary wave amplitude and velocity varying as $z^{-2/3}$, and
the width varying as $z^{1/3}$, as follows from Eqs.~(\ref{carsol}) and (\ref{kapa}).

Obviously, this approximate solution is a ring-shaped solitary wave, on top of
the cw background, which is either of the dark type (for $\alpha>0$) or of the
anti-dark type (for $\alpha<0$). Note that ring dark solitons were predicted to
occur in optical media exhibiting either Kerr \cite{ofyrds} or non-Kerr \cite{djfbam}
nonlinearities, and were later observed in experiments \cite{bou}. On the other hand,
ring anti-dark solitons were only predicted to occur in non-Kerr -- e.g., saturable
media \cite{djfbam,hec}. This picture is complemented by our analysis, according to
which a relatively strong [i.e., $d>(q/2|u_0|)^2$] nonlocal nonlinearity
can also support ring anti-dark solitary waves.

\begin{figure}[tbp]
\centering
\includegraphics
%[height=5.5cm]
[scale=0.37]{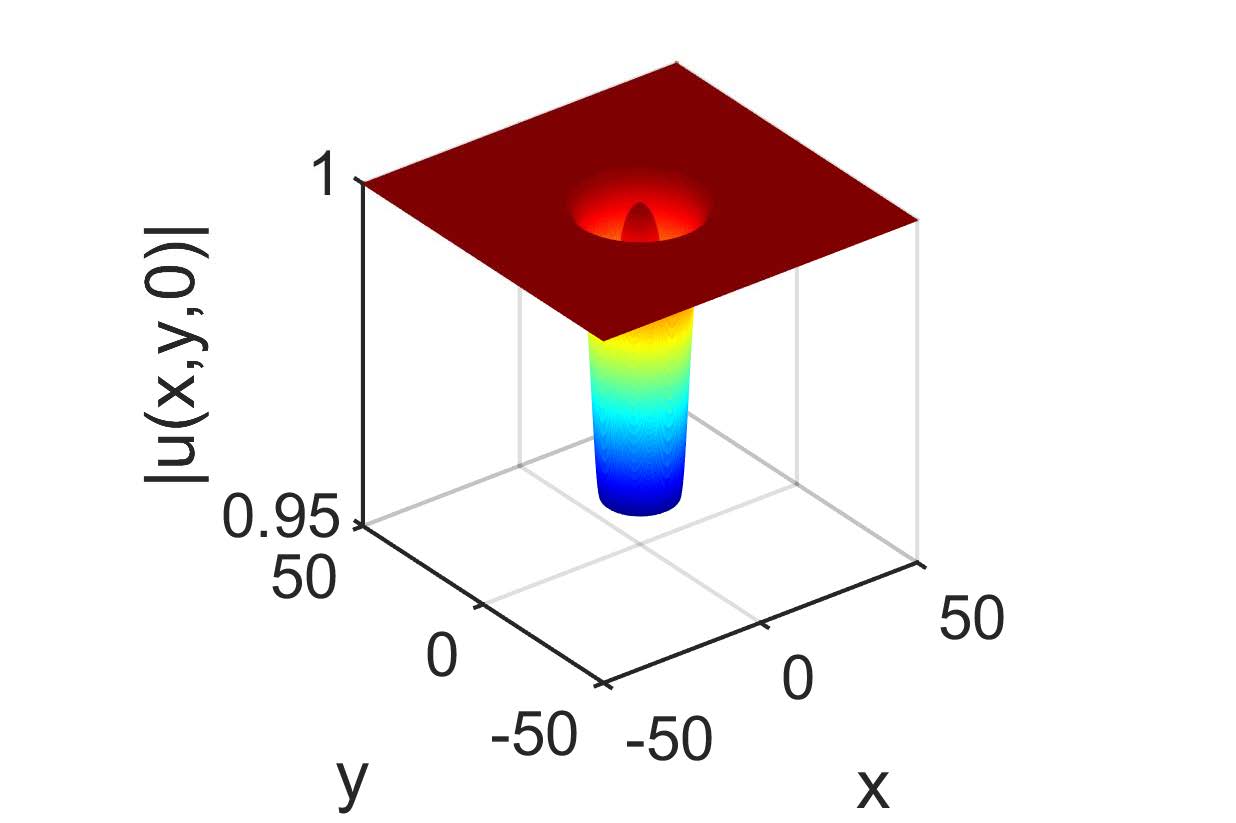}
\includegraphics
%[height=5.5cm]
[scale=0.37]{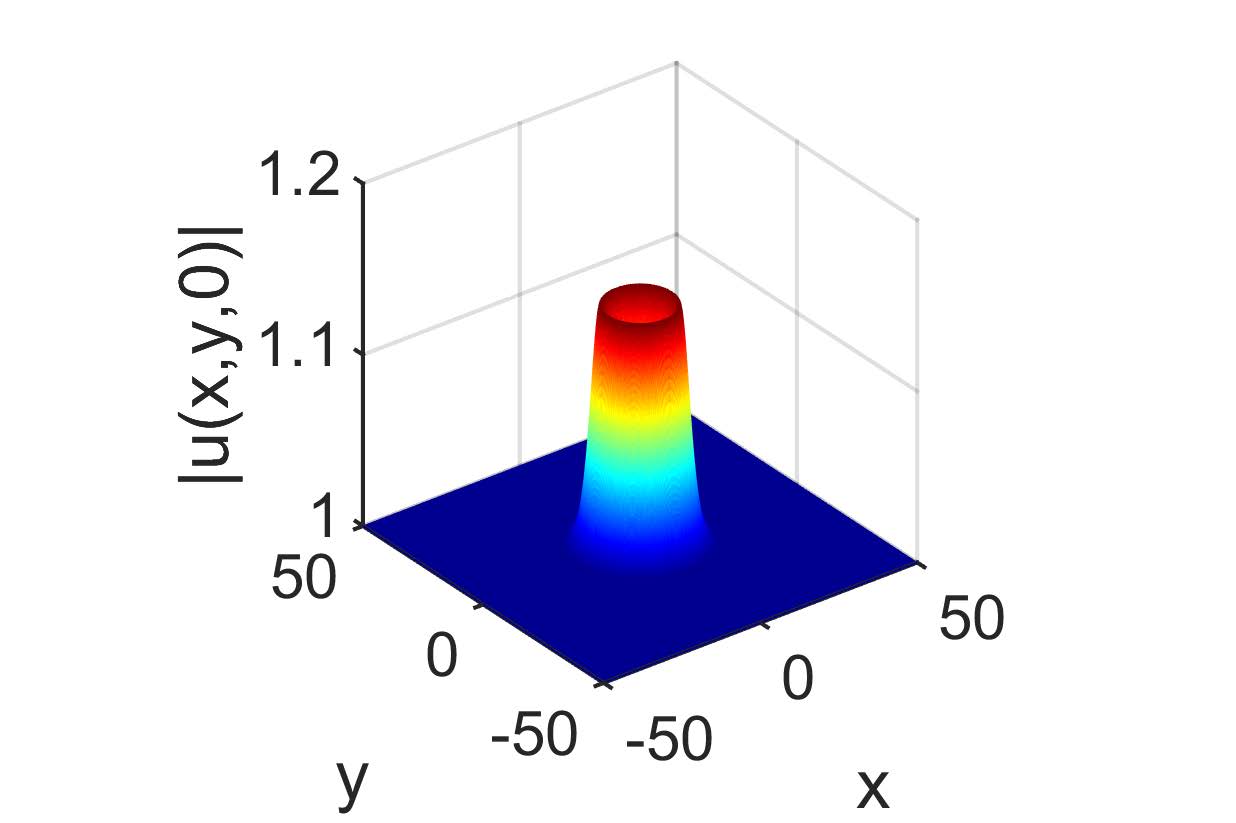}
\caption{(Color Online) Typical ring dark (left) and anti-dark soliton profiles, at $z=0$.
Both solitons have an initial radius $r_0=10$, while other parameter values are as in the
Cartesian case.}
\label{rings}
\end{figure}

\begin{figure}[ht]
\centering
\includegraphics[height=4cm]{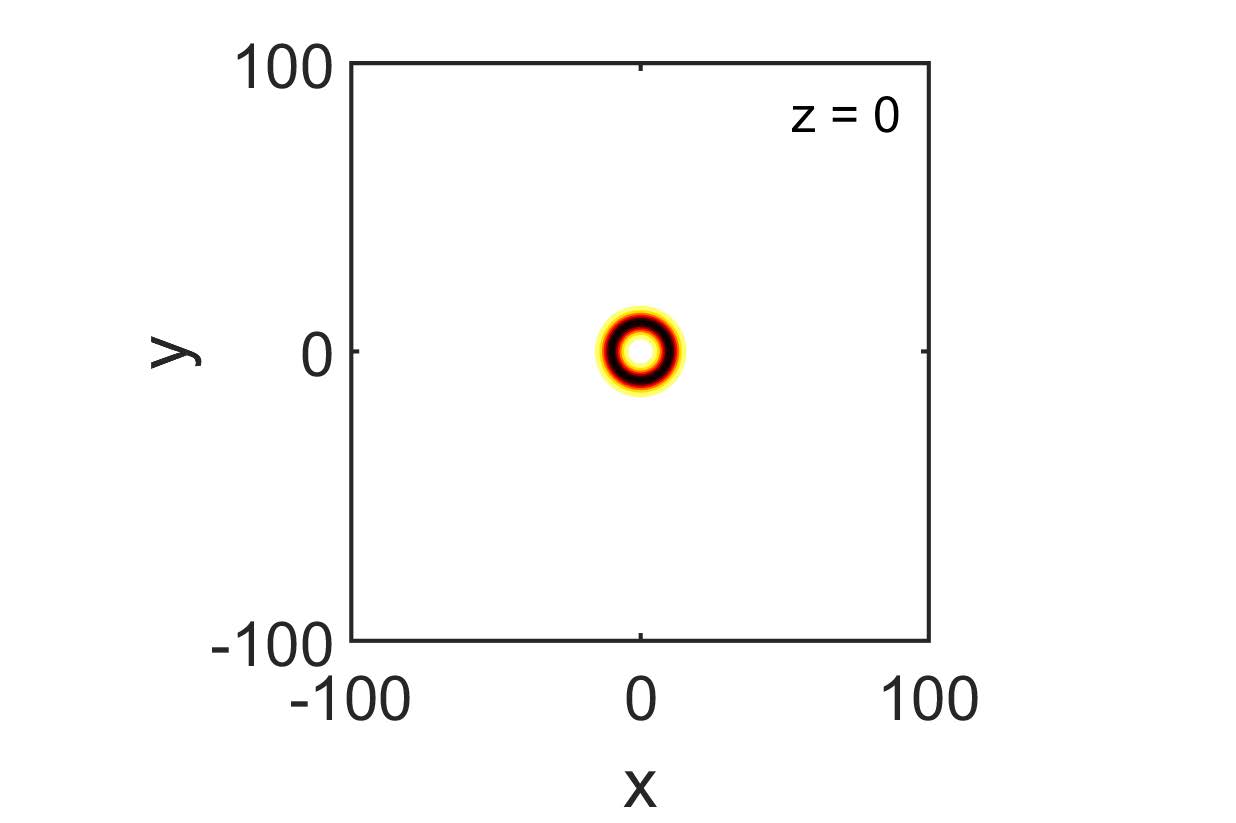}\hspace*{-1.2cm}
\includegraphics[height=4cm]{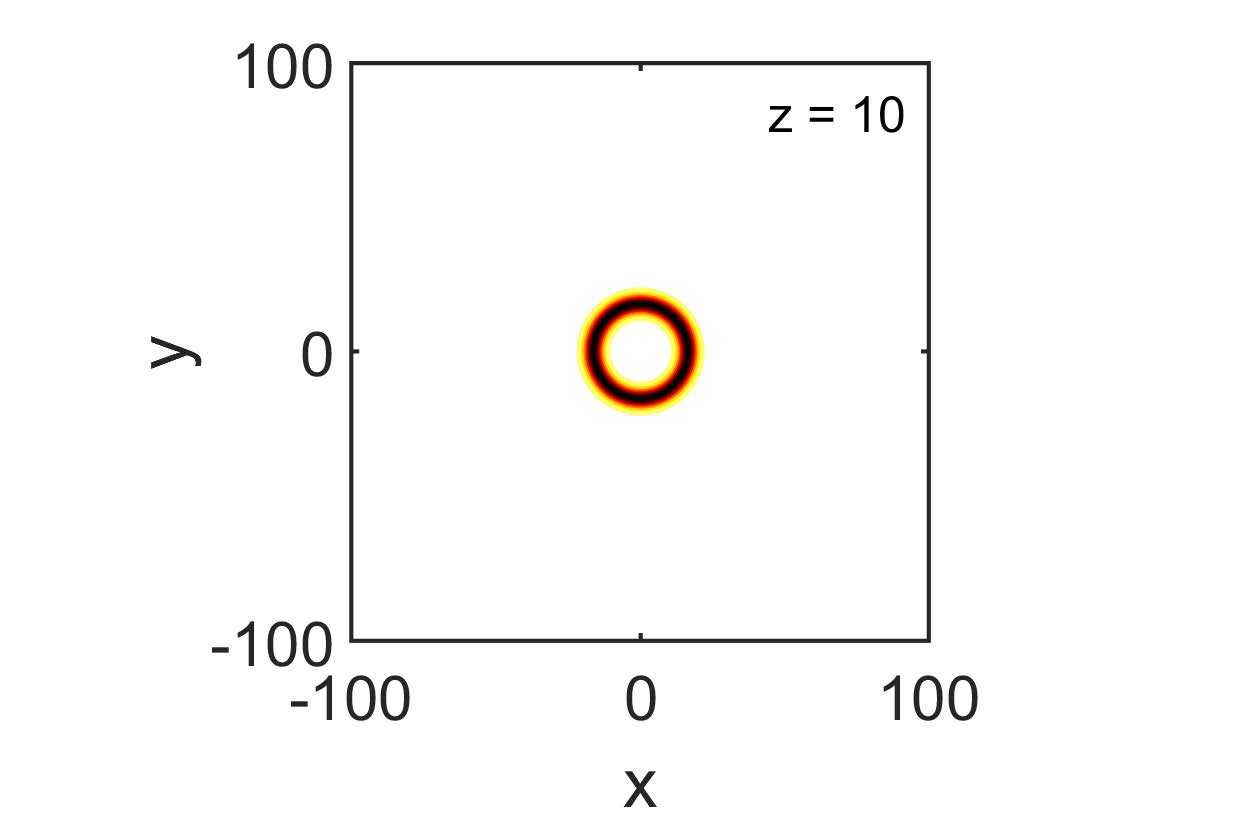}\hspace*{-1.2cm}
\includegraphics[height=4cm]{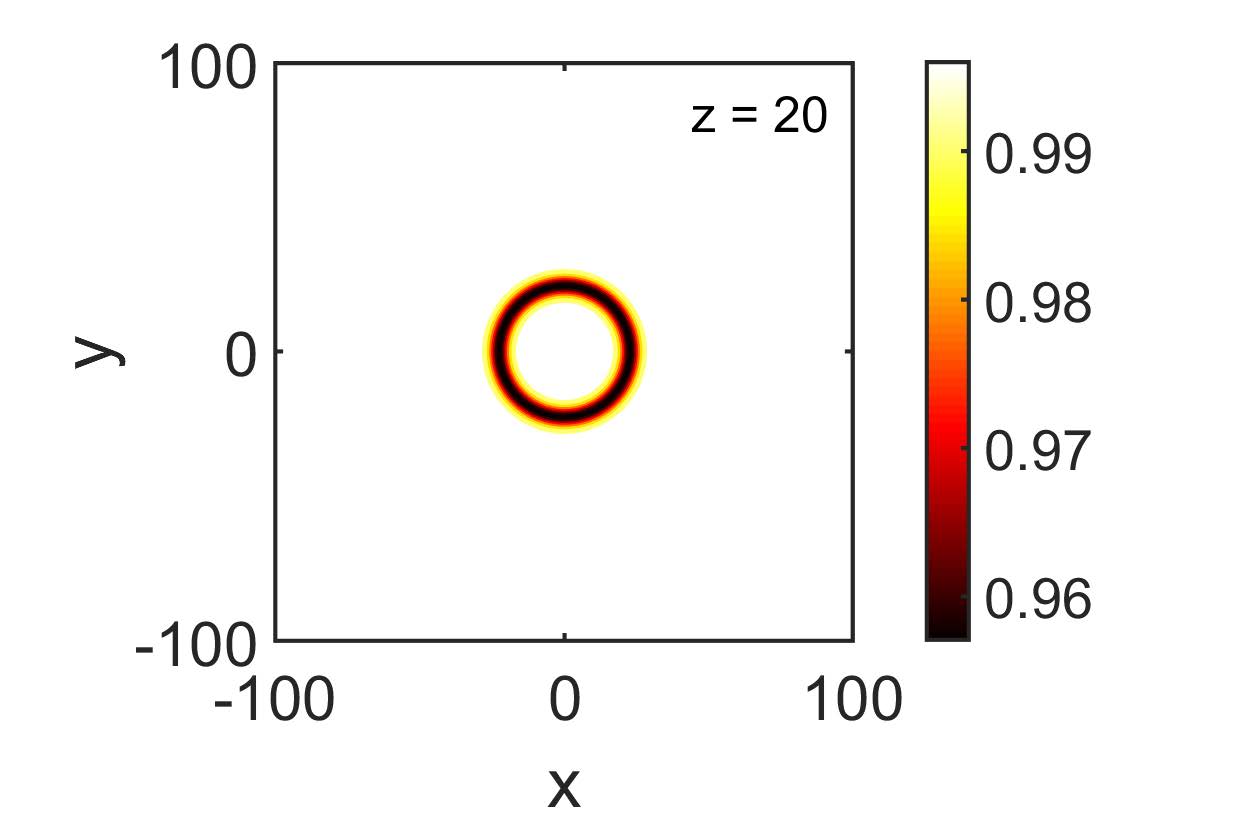}\\
\includegraphics[height=4cm]{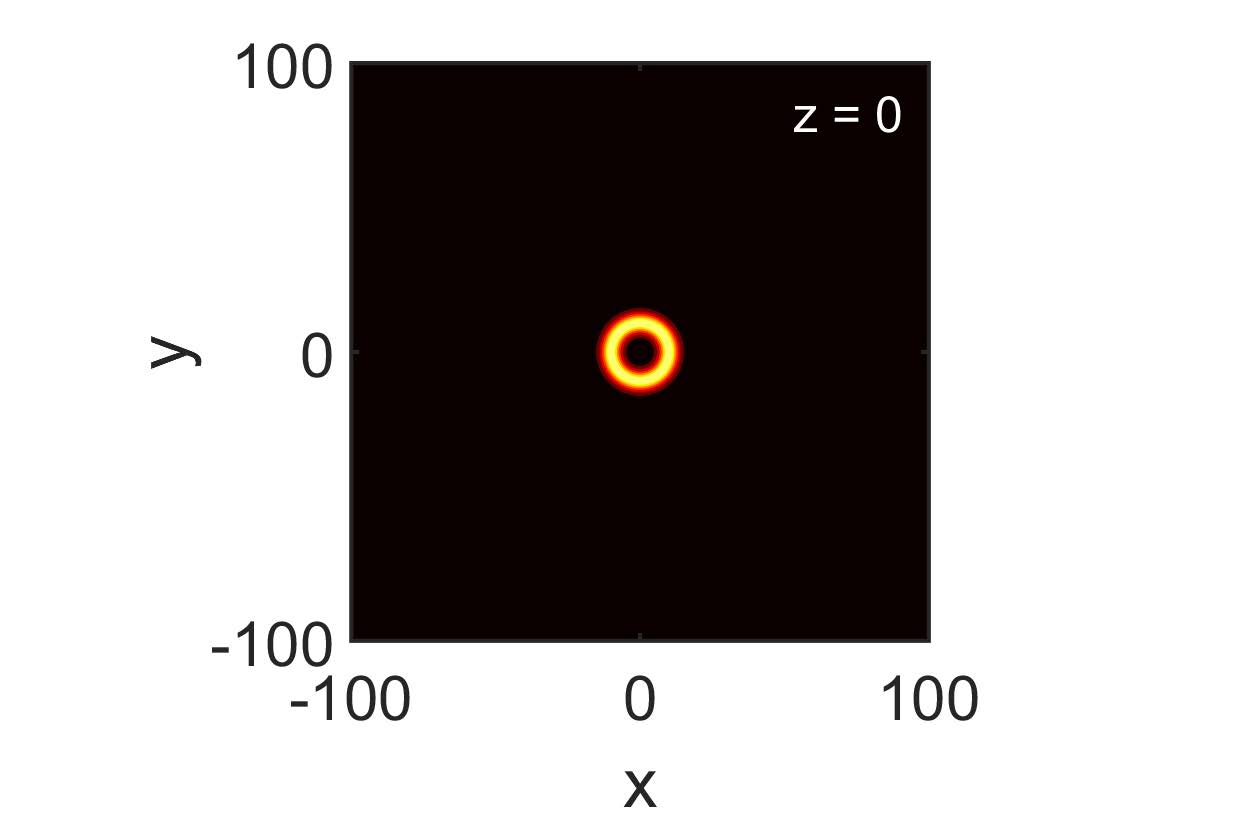}\hspace*{-1.2cm}
\includegraphics[height=4cm]{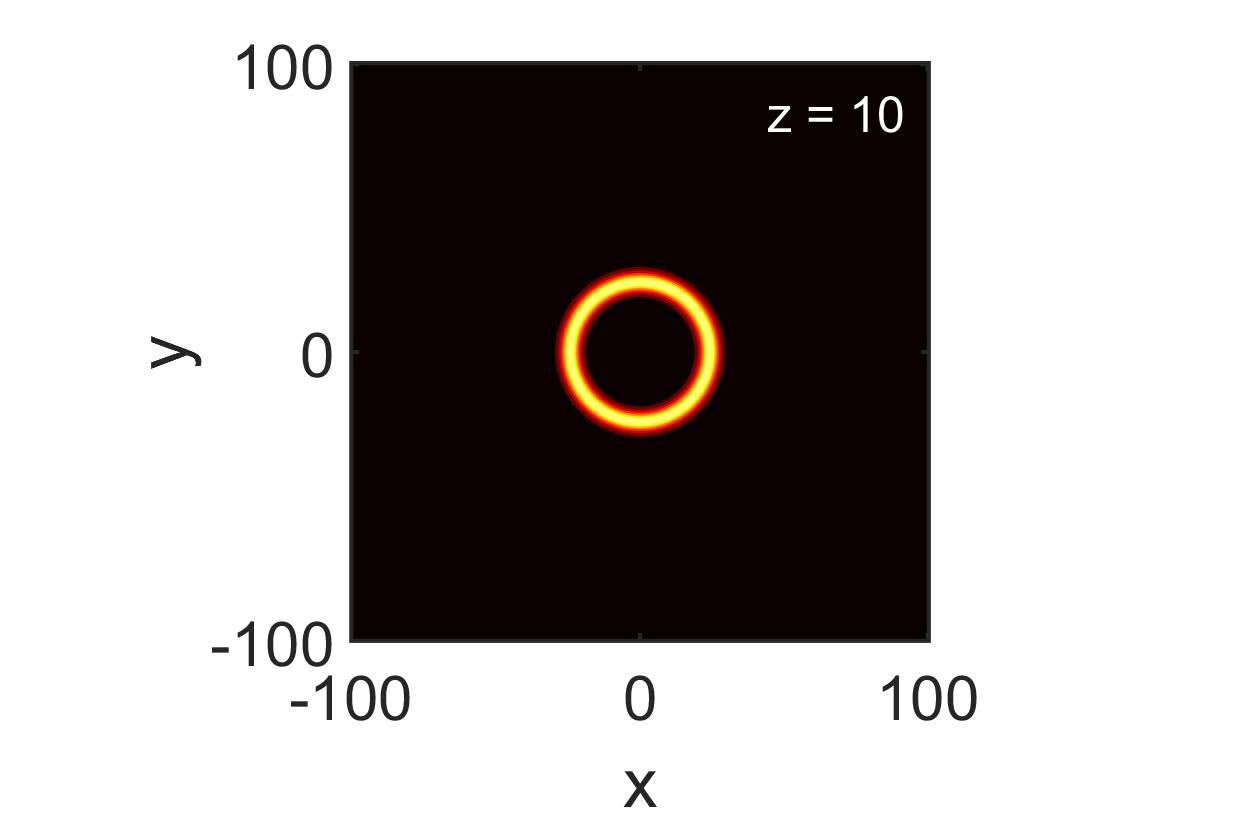}\hspace*{-1.2cm}
\includegraphics[height=4cm]{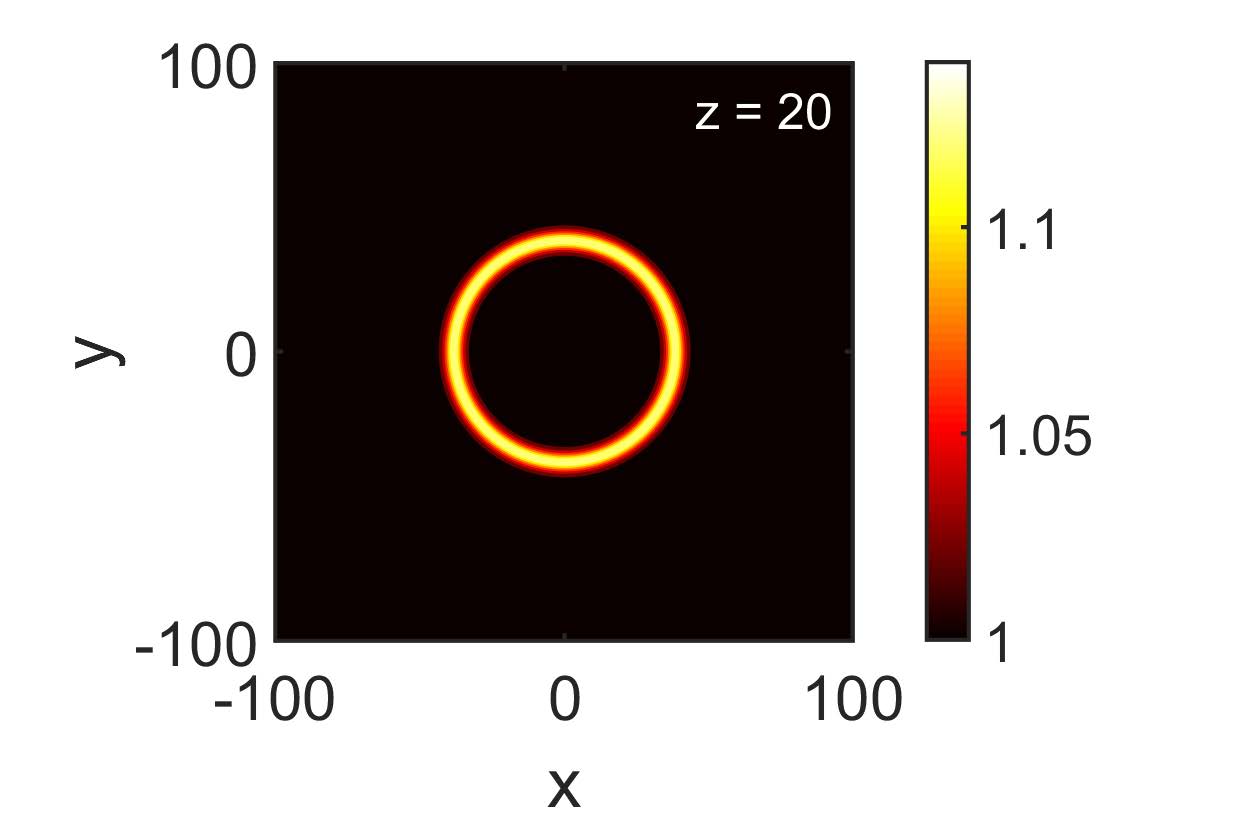}
\caption{(Color Online) Contour plots showing the evolution of the ring dark (top) and anti-dark
(bottom) ring solitons of Fig.~\ref{rings}. Results have been obtained from direct
numerical integration of Eqs.~(\ref{NLS1})-(\ref{NLS2}).}
\label{rings_evol}
\end{figure}

%\tph{
%As in the cartesian case, we depict,
In Fig.~\ref{rings}, typical ring dark and anti-dark soliton profiles, with parameter
values identical to those used in the Cartesian case, are shown at $z=0$;
%with the same parameters %as above.
both solitons have an initial radius of $r_0=10$. In addition, in Fig.~\ref{rings_evol},
contour plots depicting the evolution of the solitons' densities are shown;
these results, as before, have been obtained via direct numerical integration of
Eqs.~(\ref{NLS1})-(\ref{NLS2}). Much like the Cartesian case, the solitons
propagate undistorted, i.e., the initial rings expand outwards, keeping their shapes
during the evolution -- at least for relatively short propagation distances (see below).
It is also observed that
the solitons expand (propagate) with constant speed, with the anti-dark
soliton expanding faster than the dark soliton: indeed, the anti-dark soliton's
radius is larger than that of the dark one, at the same propagation distance.

%}

%\tph{Much like the cartesian case, the solitons propagate with constant speed, with the anti-dark
%soliton expanding/propagating faster than the dark soliton.}

%In addition,
As in the Cartesian case, the effective equation~(\ref{usJ}) can also be used to predict
(in)stability of the ring dark or anti-dark solitary waves in the $(2+1)$-dimensional setting.
In particular, and similarly to the case of the planar solitons of Eq.~(\ref{usKP}), the
case of $\alpha<0$ ($\alpha>0$), where ring dark (anti-dark) solitary waves exist, corresponds
to a KP-I (KP-II) type model. It is, thus, clear that ring dark solitary waves are expected
to be unstable, while ring anti-dark ones are predicted be stable.

Finally, let us briefly discuss the case of temporal solitary waves described by
Eq.~(\ref{KPte}).
In the $(1+1)$-dimensional setting, the underlying KdV equation has a soliton solution
similar to that in Eq.~(\ref{carsol}), with the obvious changes $\chi \rightarrow \tau$ and
$\chi_0 \rightarrow \tau_0$. However, when expressed in terms of the original fields
and variables of Eqs.~(\ref{NLS1})-(\ref{NLS2}), it is clear that the corresponding
approximate solitary wave solution is solely of the dark type: this is due to the
fact that parameter $\alpha$ is not involved in the normalization of the field $Q$
[cf. Eq.~(\ref{toKPt1})]. For the same reason, as was also mentioned in the previous
section, the higher-dimensional versions of Eq.~(\ref{KPte}) are solely of the KP-I type.
As a result, in the Cartesian $(2+1)$-dimensional setting corresponding to the usual
KP-I model, one expects the existence of stable dark ``lump'' solitary wave solutions of
the original model; these, however,
are unstable in the full 3D setting \cite{kuz1}. Finally, regarding the
cylindrical geometry, to the best of our knowledge, two-dimensional soliton solutions
of the CII model are not known.

\section{Conclusion}

In conclusion, using multiscale expansion methods, we derived asymptotic reductions of
$(3+1)$-dimensional nonlocal NLS equations, that are used to describe nonlinear waves in nematic
liquid crystals and media with a thermal nonlinearity. Working on the hydrodynamic form of the
model, both in Cartesian and cylindrical geometries, first we derived, at an intermediate stage of
the asymptotic analysis, a 3D Boussinesq equation. Then, we considered two cases, corresponding to
spatial or temporal structures and, upon introducing relevant scales and asymptotic expansions, we
reduced the Boussinesq model to KP-type equations that govern right- and left-propagating waves.
These models include various integrable and non-integrable equations at different dimensionalities
and geometries, such as the KdV and the cKdV equation, the KP-I and KP-II equations, Johnson's
equation, as well as the CI and CII equations.

We also employed these models to construct approximate solitary wave solutions of the original
nonlocal NLS model. Note that a complete discussion of the various solutions and their properties
will be provided in a later communication. As such, useful results were deduced on the type and the
stability of lower-dimensional solitary waves in higher-dimensional settings. In that regard, we
identified parameter regimes, corresponding to relatively weak or strong nonlocality, for which we
predicted the existence and stability of various solitary waves. Thus, we predicted the existence
of spatial, planar or cylindrical (ring-shaped), dark or anti-dark solitary waves, for weak or
strong nonlocality, respectively, and that dark (anti-dark) solitary waves are unstable (stable) in
the $(2+1)$-dimensional setting. Furthermore, our analysis suggested the existence of temporal
solitary waves, which become unstable in higher dimensions. Regarding approximate two-dimensional
solitary wave solutions, it was found that they may exist in the form of algebraically decaying
dark ``lumps'', which satisfy effective KP-I models; such structures may be either of the spatial
or temporal type and are supported in the weak nonlocality regime.

Our analytical predictions were also corroborated by results of direct numerical simulations.
Indeed, we have used the analytical forms of the spatial soliton profiles, in both the Cartesian
and the cylindrical geometry, we have studied their evolution stemming from the direct
numerical integration of the original nonlocal model. We have thus found that both the dark
and anti-dark soliton stripes and rings propagate undistorted, as per the effective KP picture,
at least for short propagation distances. Notice that, even for longer propagation distances,
instabilities were not observed in our simulations, which suggests that the solitons presented
here have a good chance to be observed in experiments.

Our analysis suggests various interesting directions for future studies.
For instance, it would be relevant to extend our considerations to
nonlocal models with a higher number of components (see, e.g., \cite{highnumber}). In
that case, it would be important to identify vector solitary waves in these models,
such as dark-dark and dark-bright, extending previous studies in media with
a defocusing local nonlinearity \cite{siam}.

\end{document}